\documentclass[%
  aip,
 amsmath,amssymb,
preprint,%
]{revtex4-1}
\usepackage{graphicx}% Include figure files
\usepackage{dcolumn}% Align table columns on decimal point
\usepackage{bm}% bold math
\usepackage{color}
\usepackage{float}
\usepackage{soul}
\usepackage[utf8]{inputenc}
\usepackage[T1]{fontenc}
\usepackage{mathptmx}
\usepackage{subfig}
\usepackage[inline]{enumitem}
\usepackage{caption}
\newcommand{\bogus}[1]{{}}
\begin{document}

%\preprint{AIP/123-QED}

%\title{Electron heat flux and propagating fronts in plasma thermal quench}
\title{Electron heat flux and propagating fronts in plasma thermal quench via ambipolar transport}

\author{Yanzeng Zhang}%
\email{yzengzhang@lanl.gov}
\affiliation{Theoretical Division, Los Alamos National Laboratory, Los Alamos, New Mexico 87545, USA}
\author{Jun Li}%
\affiliation{Theoretical Division, Los Alamos National Laboratory, Los Alamos, New Mexico 87545, USA}
\affiliation{School of Nuclear Science and Technology, University of Science and Technology of China, Hefei, Anhui 230026, China}
\author{Xian-Zhu Tang}%
\affiliation{Theoretical Division, Los Alamos National Laboratory, Los Alamos, New Mexico 87545, USA}


\begin{abstract}

The thermal collapse of a nearly collisionless plasma interacting with
a cooling spot, in which the electron parallel heat flux plays an
essential role, is investigated both theoretically and numerically. We
show that such thermal collapse, which is known as thermal quench in
tokamaks, comes about in the form of propagating fronts, originating
from the cooling spot, along the magnetic field lines. The slow
fronts, propagating with local ion sound speed, limit the aggressive
cooling of plasma, which is accompanied by a plasma cooling flow
toward the cooling spot. The extraordinary physics underlying such a
cooling flow is that the fundamental constraint of ambipolar transport
along the field line limits the spatial gradient of electron thermal
conduction flux to the much weaker convective scaling, as opposed to
the free-streaming scaling, so that a large electron temperature and
hence pressure gradient can be sustained. The last ion front for a
radiative cooling spot is a shock front where cold but flowing ions
meet the hot ions.

\end{abstract}

\maketitle

\section{\label{sec:level1}Introduction}

When the magnetic field lines suddenly intercept solid surfaces that
provide a sink for plasma energy and sometimes particles, a magnetized
fusion-grade plasma can undergo a thermal collapse. This can happen,
for example, in the thermal quench (TQ) of fusion plasma during either
naturally occurring or intentionally triggered, mitigated tokamak
disruptions. The naturally occurring tokamak disruption can be
triggered when large-scale MHD activities turn nested flux surfaces
into globally stochastic field lines that connect fusion-grade core
plasma directly to the divertor/first
wall~\cite{bondeson1991mhd,riccardo2002disruption,nardon2016progress,sweeney2018relationship}. As
a result, the substantial thermal energy of plasma is released within
a few
milliseconds~\cite{riccardo2005timescale,shimada2007chapter,nedospasov2008thermal},
causing severe damage to the plasma facing
components~\cite{loarte2007power} (PFCs). Major disruptions are also
intentionally triggered for disruption mitigation by injecting high-Z
impurities, for example in the form of deliberately injected solid
pellets~\cite{federici2003key,baylor2009pellet,combs2010alternative,paz2020runaway}. The
injected high-Z impurities into a pre-disruption plasma are intended
to have the hot plasma deposit its thermal energy onto the pellet, so
pellet materials can be ablated and ionized to be assimilated within
the flux surface. The ablated pellet materials thus initially forms a
radiative cooling mass (RCM) that provide strongly localized radiative
cooling for the thermal energy of the surrounding fusion-grade
plasma. The fact that the background plasma undergoes a thermal quench
by transporting energy into the RCM, which through radiation can
spread the heat load over the entire first wall, is the logic behind
this approach for thermal quench mitigation.  In both situations, the
plasma will attach to an energy sink, being a vapor-shielded wall or
the ablated pellet, and lose its energy via the fast parallel
transport to the energy sink. It must be emphasized that for
fusion-grade plasmas, which are nearly collisionless provided that the
plasma mean-free-path, $\lambda_{mfp}\sim 10km$, is much longer than
the magnetic connection length $L_B$ or the tokamak major radius, the
TQ due to the presence of a cooling spot (energy sink) is in an exotic
kinetic regime, in which the fast parallel transport along the field
lines is expected to be the dominant mechanism.  The normal
expectation is that such nearly collisionless parallel transport of
plasma thermal energy in the short magnetic connection length regime
represents a worst-case scenario of a TQ in tokamaks, where the plasma
thermal energy is released in the shortest possible time. Therefore,
understanding the plasma TQ in such an exotic regime is critical for
disruption mitigation.

For a nearly collisionless plasma, the prevailing view on the plasma
TQ is that the electron thermal conduction flux along the
magnetic field line $q_{e\parallel}$ plays the dominant role in the
heat transport. Instead of following the Braginskii
formula~\cite{braginskii}, $q_{e\parallel}$ in the collisionless limit
is considered to be constrained by the free-streaming flux
limiting~\cite{atzeni_book_2004}
\begin{equation} 
q_{e\parallel} \approx \alpha_e n_e v_{th,e} T_0, \label{eq:qe-flux-limiter} 
\end{equation} 
where $\alpha_e$ is a numerical factor~\cite{Bell-pof-1985} $\sim 0.1$
and $v_{th,e}=\sqrt{T_0/m_e}$ the electron thermal speed with $T_0$
the surrounding plasma temperature. If the flux-limiting form is
straightforwardly applied, such large $q_{e\parallel}$ will suggest a
fast TQ occurring at the electron transit time
$\tau_{tr}^e=L_B/v_{th,e}$. For a fusion-grade plasma with temperature
$\sim 10keV$, we have $\tau_{tr}^e\sim 2.5\mu s\times (L_B/100m)$,
which predicts remarkable fast TQ for $L_B<10^4m$. Such free-streaming form of thermal conduction is believed to dominate over the convective electron energy flux $\sim n_eV_{e\parallel}T_{e\parallel}$, since ambipolarity constrains $V_{e\parallel}\approx V_{i\parallel}\ll v_{th,e}$ and $V_{i\parallel}$ is bounded by the ion sound speed $c_s$.

However, it is worth noting that the inhibition of the electron thermal
conduction in the nearly collisionless plasma has been extensively
studied in astrophysics by considering the tangled magnetic
fields~\cite{tribble-mnras-1989,chandran-cowley-prl-1998} and plasma
instabilities~\cite{Jafelice-aj-1992,Balbus-Reynolds-apj-2008,Roberg-Clark-etal-prl-2018}
in order to reach the convection-dominated scenario of the thermal
conduction, which would naturally yield the {\em cooling flow} that
aggregates masses onto the cooling spot in clusters of galaxies~\cite{Fabian-ARAA-1994,Peterson-Fabian-PR-2006,Hitomi-collaboration-nature-2016,Zhuravleva-etal-nature-2014}. In a
related vein for the fusion plasma, it is well known that the
convective scaling of the electron thermal conduction is obtained at
the entrance of the steady-state sheath, in which the plasma is nearly
collisionless, as a result of the ambipolar
transport~\cite{stangeby2000plasma,stangeby1984plasma,tang2016kinetic}. In
this paper, we will show that the convective scaling with the parallel
ion flow, in the electron thermal conduction itself or its spatial
gradient, can be established throughout the bulk, quasineutral, and
nearly collisionless plasma away from the wall due to ambipolar
constraint.  Such convective scaling comes about because the cooling
of the surrounding plasma takes the form of propagating fronts that
originate from the cooling spot and the fundamental ambipolar
transport constraint between the slow fronts modifies the thermal
conduction heat flux. As a result, a robust plasma cooling flow into
the cooling spot will be developed due to the retained large electron
temperature and hence pressure
gradient~\cite{zhang-cooling-flow}. Such weaker convective scaling of
the electron thermal conduction, via itself or its gradient, is also
critically important for a slower TQ of the fusion plasmas by
modifying the core plasma cooling processes~\cite{LiTQphase}. This
should be contrasted with a straightforward application of the
flux-limiting form for electron parallel thermal conduction, which
would yield a much faster TQ on a time scale that is around a factor
of $\sqrt{m_i/m_e}$ faster.  Had one unwisely deployed the Braginskii
parallel thermal conducivity instead for such a nearly collisionless
plasma, an even faster TQ would be obtained in numerical simulations
if the plasma has collisional mean-free-path longer than the system
size.
  
\textcolor{red}{Here we follow the previous Letter of Ref.~\onlinecite{zhang-cooling-flow}
on the subject and present the details of numerical diagnostics, the
analyses of the propagating fronts' characteristics, and the
underlying physics, as well as the electron thermal conduction flux.
The first-principles fully kinetic simulations were performed with the}
VPIC~\cite{VPIC} code to investigate the parallel transport physics in
the TQ of a nearly collisionless plasma. A prototype one-dimensional
slab model is considered with a normalized background magnetic field,
where an initially uniform plasma with constant temperature $T_0\sim
10keV$ and density $n_0\sim 10^{19}m^{-3}$ is filled the whole
domain. The plasma is signified as semi-infinite with the right
simulation boundary simply reflecting the particles. We notice that
such boundary condition would not affect the plasma dynamics as long
as the later-defined electron fronts haven't arrived there yet. This
indicates that we consider $t<L_B/v_{th,e}$. However, for longer time
scale, the basic physics holds, which affects the TQ
processes~\cite{LiTQphase}. A cooling spot is modeled at the left
boundary as a thermobath that mimics a radiative cooling spot, which
conserves particles by re-injecting electron-ion pairs (equal to the
ions across the boundary) with a radiatively clamped temperature
$T_w\ll T_0$. For comparison, an absorbing boundary, as a sink to both
the particles and energy that absorbs all the particles hitting the
left boundary, is also considered for simulations. We found that these
two types of cooling spots show remarkable similarities in plasma
cooling so the absorbing boundary is quite useful for understanding
the underlying physics.

\begin{figure}[hbt]
\centering
\includegraphics[width=0.45\textwidth]{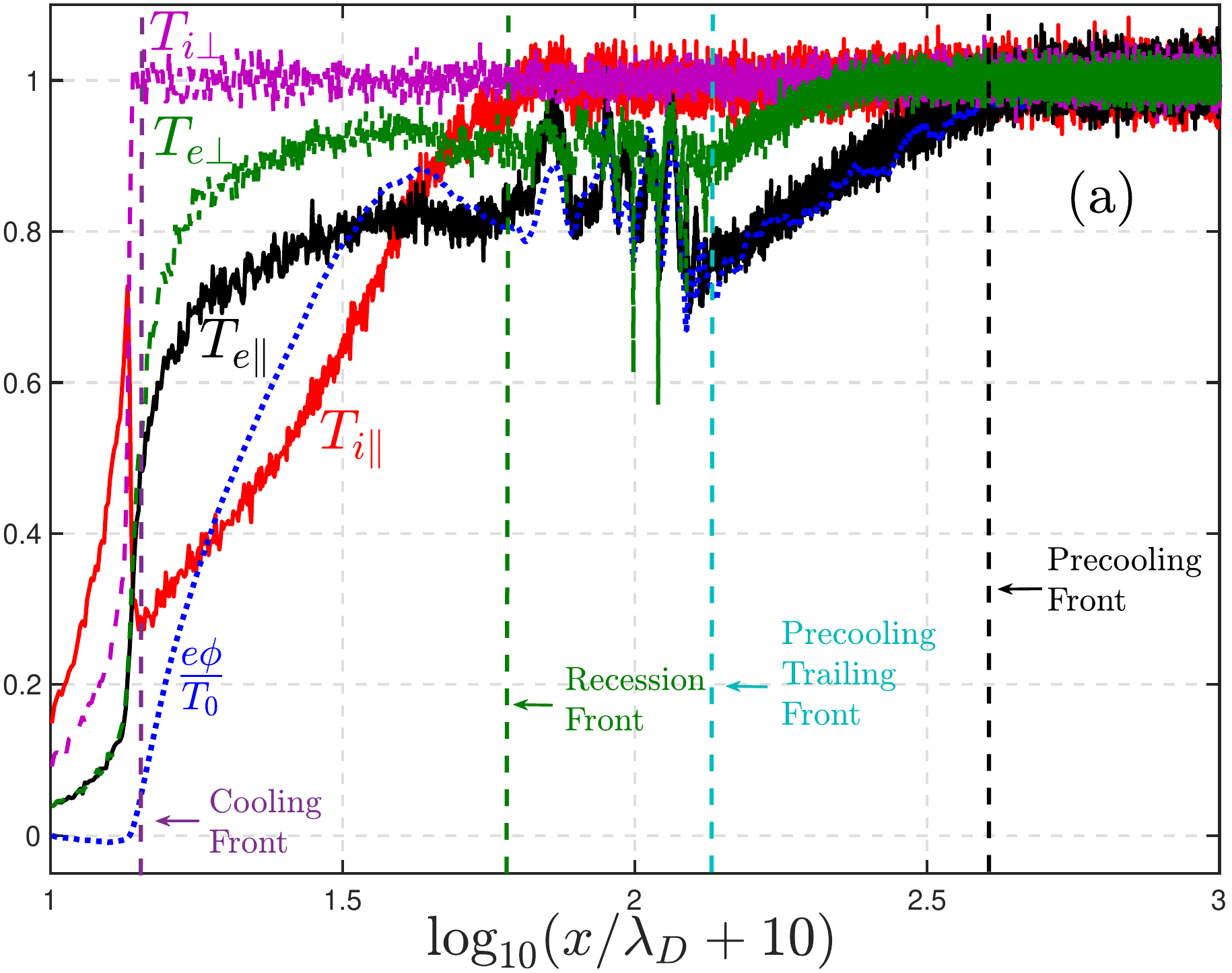}
\includegraphics[width=0.45\textwidth]{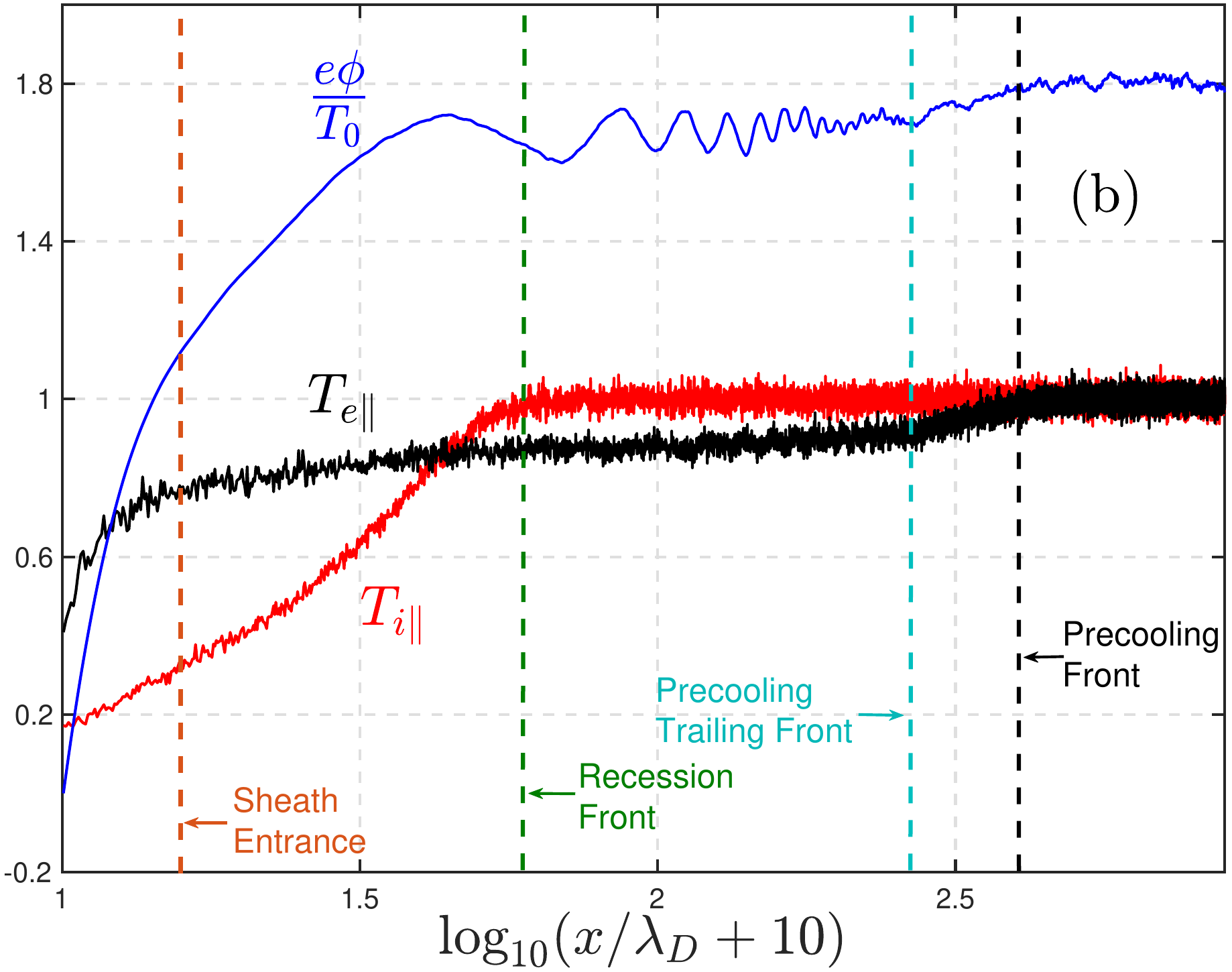}
\caption{Normalized electron and ion temperature (by $T_0$), and
  electrostatic potential (by $T_0/e$) at $\omega_{pe}t=163$ for the
  thermobath boundary with $T_w=0.01T_0$ (a) and the absorbing
  boundary (b) from the first-principles kinetic simulations of
  collisionless plasma (similar results are obtained for nearly
  collisionless plasma with $\lambda_{mfp}\gg L_B$). Notice that the
  perpendicular temperature for the absorbing boundary are nearly
  unperturbed and hence ignored. The sheath entrance here is denoted
  as where the plasma parameters remain quasi-steady. We should
  emphasize that the electrostatic potential is directly integrated
  from the instantaneous electric field that contains large amplitude
  Langmuir waves, which thus is only used as a guide for the
  qualitative behavior of the quiescent ambipolar potential. }
\label{fig:diagram}
\end{figure}

For both thermobath and absorbing boundary, the TQ is found to be
governed by the formations of propagating fronts (e.g., see
Fig.~\ref{fig:diagram}). Particularly, for the former (latter), there
are four (three) fronts: the first two have speeds scale with the
electron thermal speed $v_{th,e}$ and thus are named electron fronts,
while the other two (one) propagate at speeds that scale with the
local ion sound speed $c_s$ and thus named ion fronts. Based on the
underlying physics and their roles in the TQ dynamics, these fronts
can be named the precooling front (PF), precooling trailing front
(PTF), recession front (RF), and cooling front (CF, for the thermobath
boundary only), respectively, as illustrated in
Fig.~\ref{fig:diagram}. The precooling and recession fronts describe
the onset of $T_\parallel$ cooling of electrons and ions,
respectively, which are independent of the cooling spot types. In
contrast, the precooling trailing and cooling fronts will not only
play a role in $T_\parallel$ cooling but also reflect the onset of
$T_\perp$ cooling via the dilution with the cold recycled particles
for the thermobath boundary.  \textcolor{red}{It should be noted that the propagating
fronts for the rapid cooling of a nearly collisionless plasma, as
described and reported here, are not the artifact of the boundary
conditions deployed in the simulation.  In a forthcoming paper that
focuses on the impurity ion assimilation by a cooling plasma, the same
front propagation physics are found in a plasma where a hot and dilute
plasma cools against a cold and dense plasma, that is initially in
pressure balance between the two regions.}

\iffalse which, in the absence of pitch-angle
scattering induced by the collisions and wave-particle
interaction~\cite{guo2012ambipolar,zhangwhistler}, can be only cooled
via the dilution with the cold recycled particles.  \fi

The rest of the paper is organized as follows. In section
\ref{sec:electronfronts} we elucidate the underlying physics of
electron fronts, while the ion front(s) physics are investigated in
section \ref{sec:ion-front}. The electron thermal conduction flux
within the recession layer (the region between the recession and
cooling front for the thermobath boundary or between the recession
front and sheath entrance for the absorbing boundary), which is
essential for the formation of the plasma cooling flow, will be
discussed in section \ref{sec:heatflux}. We will conclude in section
\ref{sec:conclusion}.

\section{\label{sec:electronfronts}Electron fronts}

\begin{figure}[hbt]
\centering
\includegraphics[width=0.45\textwidth]{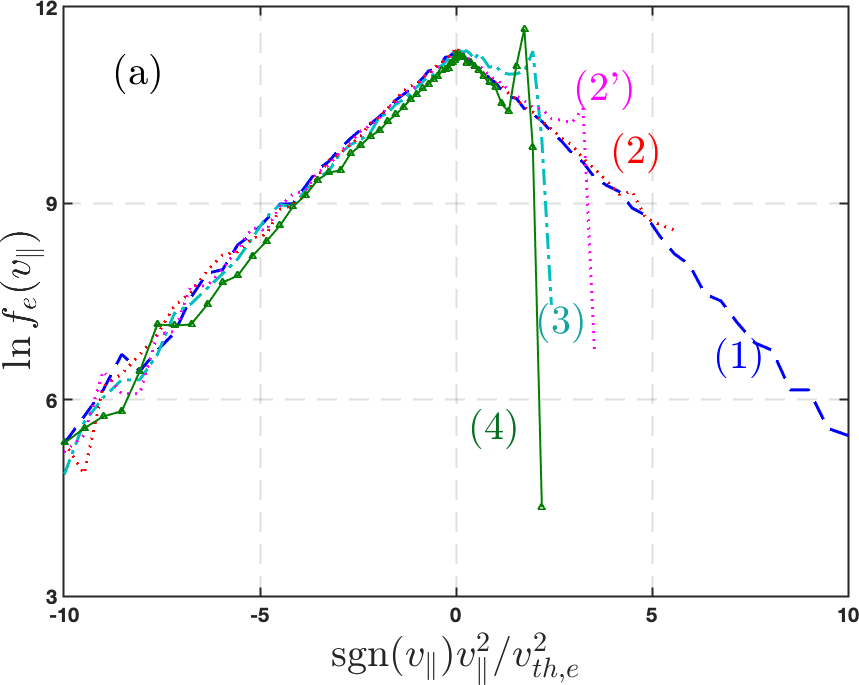}
\includegraphics[width=0.45\textwidth]{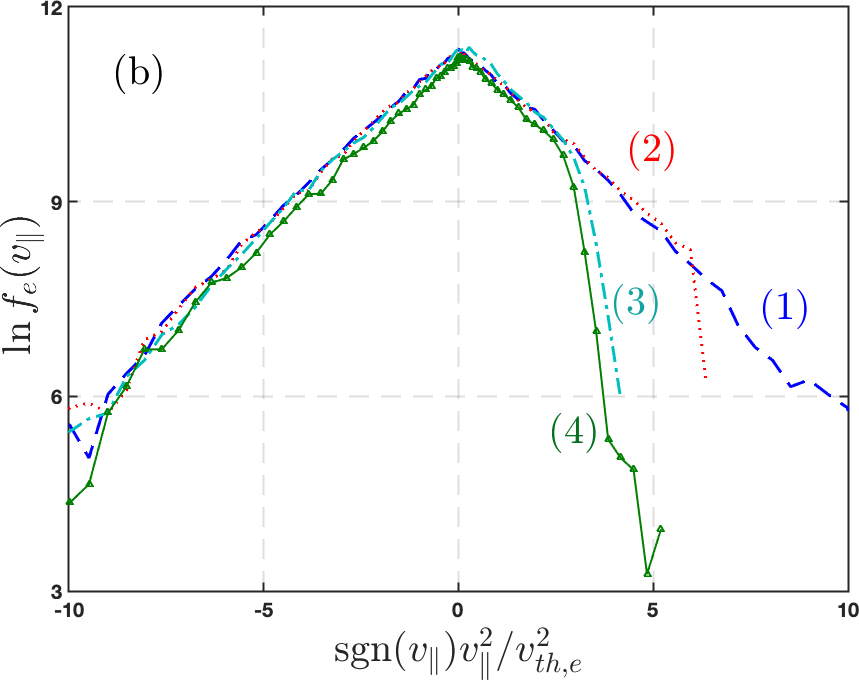}
\caption{Electron distributions at different locations from VPIC
  simulations corresponding to Fig.~\ref{fig:diagram}. (a) is for the
  thermobath boundary and (b) is for the absorbing boundary. These
  distributions are chose from: (1) ahead of the precooling front, (2)
  between the precooling and precooling trailing fronts (for the
  thermobath boundary, it is ahead of the cold electron beam front
  where $T_{e\perp}$ is unperturbed), (3) between the recession and
  precooling trailing fronts and (4) within the recession layer. For
  the thermobath boundary, ($2'$) is behind the cold electron beam
  front but ahead of the precooling trailing front. }
\label{fig:electron-distribution}
\end{figure}

We first investigate the physics underlying the electron fronts. In a
nearly collisionless plasma, the cooling of the parallel electron
temperature $T_{e\parallel}$ is mainly due to the cutoff of electron
distribution function as a result of the loss of high energy electrons
that overcome the electrostatic potential barrier, i.e.,
$v_\parallel>v_c$ where $v_c=\sqrt{2e\Delta \Phi/m_e}$ with $\Delta
\Phi=\Phi(x)-\Phi_{\rm min}$, $\Phi(x)$ being the local electrostatic
potential and $\Phi_{\rm min}$ the minimum potential. For the
absorbing boundary, $\Phi_{\rm min}$ is the boundary potential, while
for the thermobath boundary, $\Phi_{\rm min}$ is the local minimum
potential just behind the cooling front as shown in
Fig.~\ref{fig:diagram}. It must be highlighted that, for the
thermobath boundary, the same ambipolar field can accelerate some cold
recycled electrons towards the upstream to form an electron beam with
velocity $V_b\approx v_c$ due to the energy gain at the ambipolar
potential, e.g., see Fig.~\ref{fig:electron-distribution}, the front
of which is between the electron fronts that will be discussed later
in this section. As a result, the electron distribution can be
approximated as
\begin{equation} 
f_e(v_\parallel,v_\perp) = \frac{n_m\left(\Phi(x)\right)}{\sqrt{2\pi}
  v_{th,e}^3} e^{-\left(v_\parallel^2+v_\perp^2\right)/2v_{th,e}^2}
\Theta\left(1-\frac{v_\parallel}{v_c}\right)+ \frac{n_b}{2\pi
  v_\perp}\delta(v_\perp)\delta(v_\parallel -
v_c),\label{eq:truncated-fe}
\end{equation} 
where $\Theta(x)$ is the Heaviside step function that vanishes for
$x<0$ and $\delta(x)$ the Dirac delta function, $n_m$ is the plasma
density associated with the cutoff Maxwellian (the original
surrounding electrons) and $n_b$ the recycled (cold) electron beam
density. For the absorbing boundary (or ahead of the cold electron
beam front for the thermobath boundary), $n_b=0$, for which one
obtains
\begin{align}
n_e=&\frac{1+\textup{Erf}(v_c/\sqrt{2}v_{th,e})}{2}n_m+n_b, \label{eq:ne-coldbeam}\\
 n_eV_{e\parallel}=&-\frac{n_mv_{th,e}}{\sqrt{2\pi}} e^{-v_c^2/2v_{th,e}^2}+n_bv_c, \label{eq:nV-cold-beam}\\
 n_eT_{e\parallel}=&n_mT_0\left[\frac{1+\textup{Erf}(v_c/\sqrt{2}v_{th,e})}{2} - \frac{v_{c}}{\sqrt{2\pi}v_{th,e}}e^{-{v_c^2}/{2v_{th,e}^2}}\right] + n_bT_0\frac{v_c^2}{v_{th,e}^2} - n_e T_0\frac{V_{e\parallel}^2}{v_{th,e}^2},\label{eq:neTe-coldbeam}\\
 q_{en}=& - \frac{n_m v_{th,e} T_0}{\sqrt{2\pi}}\left(\frac{v_c^2}{v_{th,e}^2}+2\right) e^{-v_c^2/2v_{th,e}^2}+n_bT_0v_c\frac{v_c^2}{v_{th,e}^2} -3n_eT_{e\parallel}V_{e\parallel} - n_e m_e V_{e\parallel}^3.
  \label{eq-qen}
\end{align}
Here $n_e=\int f_e d^3\mathbf{v}$ is the electron density,
$V_{e\parallel}=\int v_\parallel f_e d^3\mathbf{v}/n_e$ the parallel
electron flow, $T_{e\parallel}=\int m_e \tilde{v}_\parallel^2 f_e
d^3\mathbf{v}/n_e$ the parallel electron temperature, and $q_{en} =
\int m_e \tilde{v}_\parallel^3 f_ed^3\mathbf{v}$ the parallel thermal
conduction flux of the parallel degree of freedom~\cite{Chew1956},
where $\tilde{v}_\parallel\equiv(v_\parallel-V_{e\parallel})$.

The ambipolar transport constraint implies $V_{e\parallel}\approx
V_{i\parallel}$, one immediate result of which is that the last terms
in $T_{e\parallel}$ and $q_{en}$ that are proportional to
$V_{e\parallel}^{2,3}$ are negligible as $V_{e\parallel}\approx
V_{i\parallel}\tilde{<} v_{th,i} \ll v_c\sim v_{th,e}$ with
$v_{th,i}=\sqrt{T_0/m_i}$. This is especially so between the
electron fronts where the plasma flow is nearly unperturbed since it
is controlled by the ion fronts.  Without the plasma flow ahead of the
ion fronts, the plasma energy equation shows that the collapse of
$T_{e\parallel}$ is mainly driven by the conduction heat flux $q_{en}$
\begin{align}
  n_e\left(\frac{\partial}{\partial t} T_{e\parallel} +
  V_{e\parallel}\frac{\partial}{\partial x} T_{e\parallel}\right) +
  2n_e T_{e\parallel}\frac{\partial}{\partial x}V_{e\parallel} +
  \frac{\partial}{\partial x} q_{en} = 0.\label{eq-electron-energy}
\end{align}

Between the electron fronts, the fraction of electron beam for the
thermobath boundary is small $n_b\ll n_m$ (notice that $n_b=0$ for the
absorbing boundary). Thus, the ambipolar transport constraint
$V_{e\parallel}\approx V_{i\parallel}$ requires $v_c>\sqrt{2}v_{th,e}$
as seen from Eq.~(\ref{eq:nV-cold-beam}) and
Fig.~\ref{fig:electron-distribution}. Such a condition gurantees that
\begin{equation} 
\frac{d q_{en}}{dx} \approx  n_e v_c \frac{\partial T_{e\parallel}}{\partial x}.
\end{equation} 
As a result, the energy equation in Eq.~(\ref{eq-electron-energy}) has
the solution $T_{e\parallel}=T_{e\parallel}(x-v_ct)$ for $n_e\approx
n_0$, revealing that $v_c$ is the recession speed of $T_{e\parallel}$
or the velocity space void in $f_e(v_\parallel)$ propagates upstream
with a speed of $v_c$. It must be emphasized that the electron fronts
are fast $\sim v_{th,e}$ but only produce a modest amount of
$T_{e\parallel}$ cooling as seen from Eq.~(\ref{eq:neTe-coldbeam}) for
$v_c\gg v_{th,e}$, which results from the ambipolar transport
constraint. Moreover, Eq.~(\ref{eq-qen}) illustrates that the electron
parallel conduction flux does scale as the free-streaming limit,
$q_{en} \propto n_e v_{th,e} T_0$ but with a small coefficient as a
function of $v_c(\Phi)$.

\begin{figure}[hbt]
\centering
\includegraphics[width=0.3\textwidth]{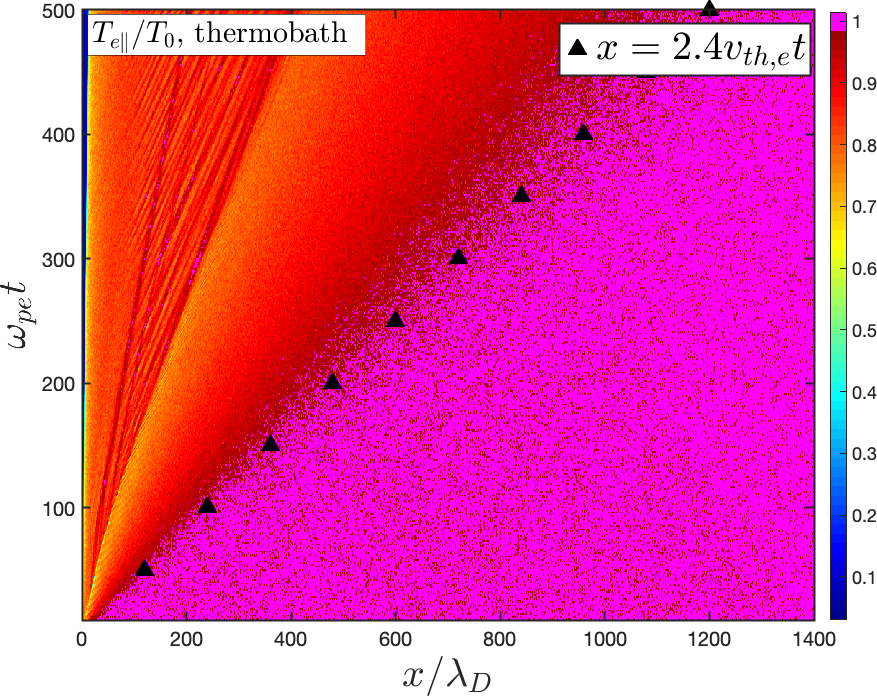}
\includegraphics[width=0.31\textwidth]{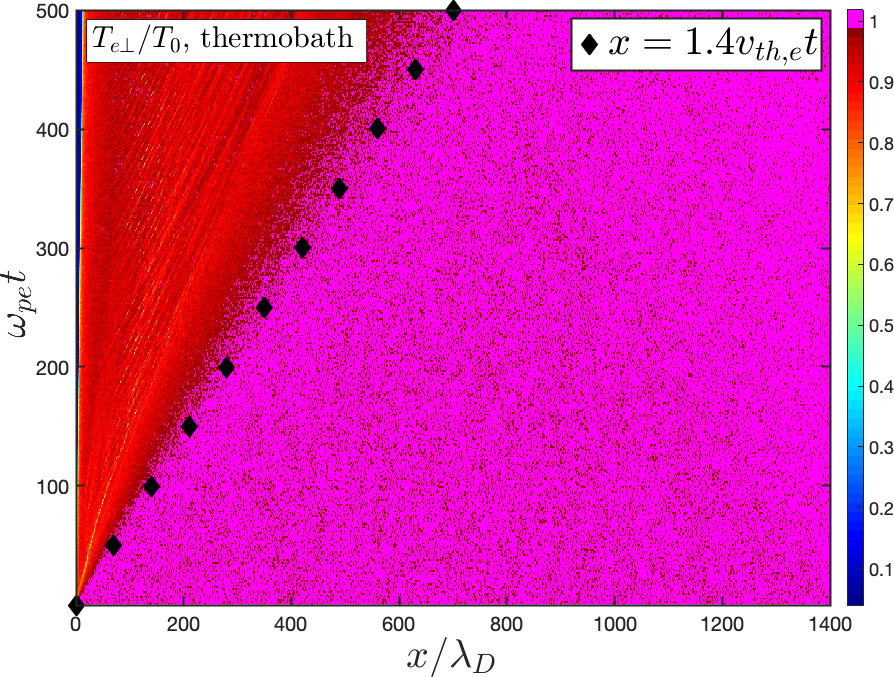}
\includegraphics[width=0.3\textwidth]{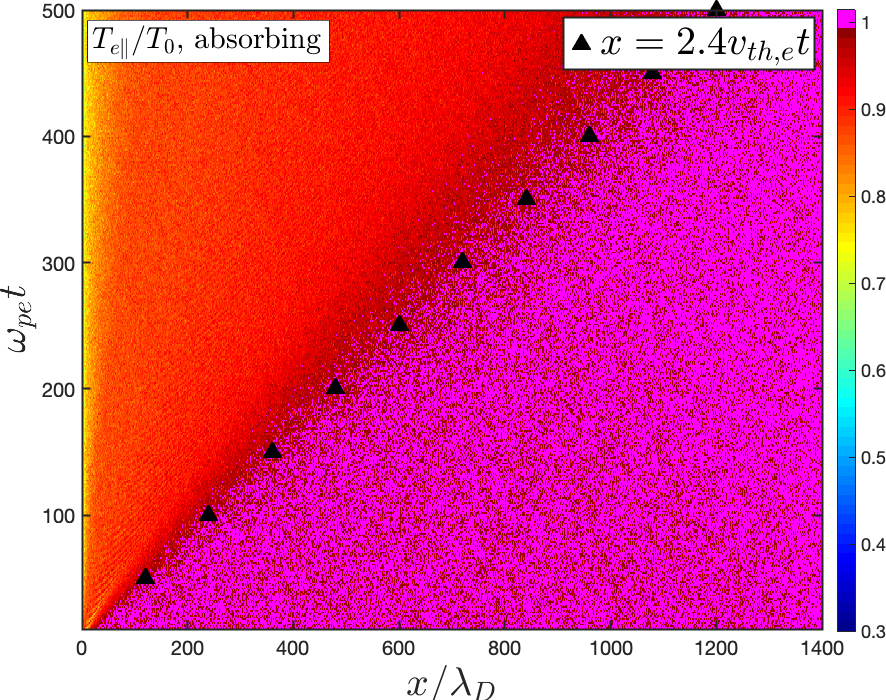}
\caption{Contour plots of normalized electron temperature and the
  corresponding fitting precooling front (black triangles) for the
  same simulations in Fig.~\ref{fig:diagram}, where the left boundary
  conditions are labeled. For the thermobath boundary, we also plot
  the fitting front of $T_{e\perp}$ collapse (black diamonds). The
  nearly unperturbed values are colored as magenta. }
\label{fig:contour-Te-reflux}
\end{figure}

Now we can define the electron fronts and provide their propagating speeds by using the local $v_c$. The precooling front (PF) is used to denote the onset of $T_{e\parallel}$ cooling, which can be defined as where $T_{e\parallel}(v_c)$ has a detectable cooling.  It is independent of the boundary condition since it is ahead of the recycled electron beam front.  Considering the VPIC noise, we choose the speed of the PF as
\begin{equation}
U_{PF}=2.4v_{th,e},   
\end{equation}
for which $T_{e\parallel}(v_c=U_{PF})\approx 0.95T_0$ as seen from Eq.~\eqref{eq:neTe-coldbeam} with $n_b=0$.  It agrees well with the simulation results in Fig.~\ref{fig:contour-Te-reflux}. While the precooling trailing front (PTF) comes about because the ambipolar potential and hence the plasma cooling mainly varies behind the recession front (e.g., see Fig.~\ref{fig:diagram}) due to the limit of the plasma flow. Therefore, there must be a smallest cutoff velocity $v_c^{\rm min}=\sqrt{2e(\Delta
  \Phi)_{\rm RF}/m_e}\sim v_{th,e}$ ahead of the recession front (RF) determined by the reflecting potential, $(\Delta\Phi)_{\rm RF}=\Phi_{\rm RF}-\Phi_{\rm min}$ with $\Phi_{\rm RF}$ the potential at the RF. Such $v_c^{\rm min}$ will define the PTF speed
\begin{equation}
    U_{PTF}=v_c^{\rm min}=\sqrt{2e(\Delta
  \Phi)_{\rm RF}/m_e}.\label{eq-U-PTF}
\end{equation}
The fact that the cutoff velocity between the PTF and RF satisfies $v_c\approx v_c^{\rm min}$ demonstrates that the PTF will leave a nearly constant $T_{e\parallel}=T_{e\parallel}(U_{PTF})$ behind. For an absorbing boundary with $n_b=0$, from Eq.~(\ref{eq:neTe-coldbeam}) we obtain
\begin{equation} 
  T_{e\parallel}(U_{PTF}) \approx T_0 \left[1- 
    \sqrt{{e\left(\Delta\Phi\right)_{\rm RF}}/{\pi T_0}} e^{- 
      {e\left(\Delta\Phi\right)_{\rm RF}}/{T_0}} \right]. \label{eq-Te-PTF}
\end{equation} 
For the thermobath boundary, the aforementioned cold electron beam,
which are accelerated by the electrostatic potential all the way to
$\sim U_{PTF}$, will reduce $T_{e\parallel}(U_{PTF})$. This is because
such electron beam will reduce the reflecting potential for the
thermobath boundary compared to that for the absorbing boundary as
shown in Fig.~\ref{fig:diagram} (this will be studied in section
\ref{sec:heatflux}), which results in a slower PTF (smaller $U_{PTF}$)
and hence deeper cooling of $T_{e\parallel}$ than those for the
absorbing boundary.

Interestingly, the faster PTF for the absorbing boundary nearly
coincides with the $T_{e\perp}$ collapse front (cold recycled electron
beam front) for the thermobath boundary as shown in
Fig.~\ref{fig:diagram}. This is because, at the beginning of the
thermal collapse, the number of recycled electrons is not high enough
to cause an appreciable reduction of the reflecting potential compared
to that for the absorbing boundary. Such nearly equalized reflecting
potential $(\Delta \Phi)_{\rm RF}$ will accelerate the recycled
electrons, which dilutely cool $T_{e\perp}$, to $U_{PTF}$ for the
absorbing boundary case. A fitting speed of $1.4v_{th,e}$ for
$T_{e\perp}$ precooling front is shown in
Fig.~\ref{fig:contour-Te-reflux}.
 
\section{\label{sec:ion-front} Ion fronts}

The ion front(s) not only describes the plasma temperature cooling but
also controls the plasma density evolution and the associated cooling
flow generation. This section is dedicated to studying the underlying
physics of the ion front(s), taking into account that the plasma
thermal conduction is forced to be convective, in the form of either
itself or its gradient, due to the ambipolar transport, which will be
investigated in the next section. As a result, the electron pressure
gradient remains large to drive plasma flow towards the cooling spot,
which is known in astrophysics as the cooling flow.~\cite{Fabian-ARAA-1994,Peterson-Fabian-PR-2006,Hitomi-collaboration-nature-2016,Zhuravleva-etal-nature-2014}

We first highlight that the cooling front (CF) for the thermobath
boundary case is the front of cold recycled ions. Therefore, the
plasma ahead the CF is free of cold recycled ions. Such a block of
cold ions by the CF is complete and thus different from that of the
cold recycled electrons discussed above, where some of them can
cross the CF and penetrate into the upstream plasma as cold recycled electron
beams. As a result, we can examine the recession front (RF) and
the recession layer between the RF and CF by ignoring the recycled ions. In physics
terms, the recession layer is a rarefaction wave in the nearly
collisionless plasma, which is different from the rarefaction wave
formed in a cold plasma interacting with a solid
surface~\cite{allen1970note,cipolla1981temporal,breizman2021plasma}
considering the large plasma temperature and pressure gradients and
the nature of the plasma heat fluxes. But the common feature of the
rarefaction wave is that the plasma parameters recede steadily with
local speed $\sim c_s$ and thus we can seek self-similar solutions of
$n_{e,i},V_{e,i\parallel}, T_{e,i\parallel}$ as functions of the
self-similar variable $\xi\equiv x/t \sim c_s$ from the minimum model
for an anisotropic
plasma~\cite{Chew1956,Chodura_1971,guo-tang-pop-2012a}
\begin{align}
\label{eq-ion-density}
& \frac{\partial }{\partial t}n_{i} + \frac{\partial}{\partial x} \left(n_{i} V_{i\parallel}\right) = 0, \\
  & m_{i} n_{i}\left(\frac{\partial }{\partial t} V_{i\parallel}+ V_{i\parallel}\frac{\partial}{\partial x} V_{i\parallel} \right)
  + \frac{\partial }{\partial x} (p_{i\parallel}+p_{e\parallel}) = 0,\label{eq-ion-momentum} \\
& n_i \left( \frac{\partial}{\partial t} T_{i\parallel} +V_{i\parallel}  \frac{\partial}{\partial x} T_{i\parallel}\right)+ 2n_iT_{i\parallel}\frac{\partial}{\partial x}V_{i\parallel} +\frac{\partial}{\partial x} q_{in}
  = 0.\label{eq-ion-temperature}
\end{align}
The force balance of electrons $e n_{e} E_\parallel\approx - \partial p_{e\parallel}/\partial x$ and the quasi-neutral condition $n_e=Zn_i$ are used with $Z$ being the ion charge number, and $p_{e,i\parallel} = n_{e,i}T_{e,i\parallel}$. Notice that the quasi-neutrality ensures $V_{e\parallel}\approx V_{i\parallel} \ll v_{th,e}$ in the absence of net current and thus the force balance of electrons is valid within the recession layer due to the small inertia of electrons.

\begin{figure}[hbt]
\centering
\includegraphics[width=0.3\textwidth]{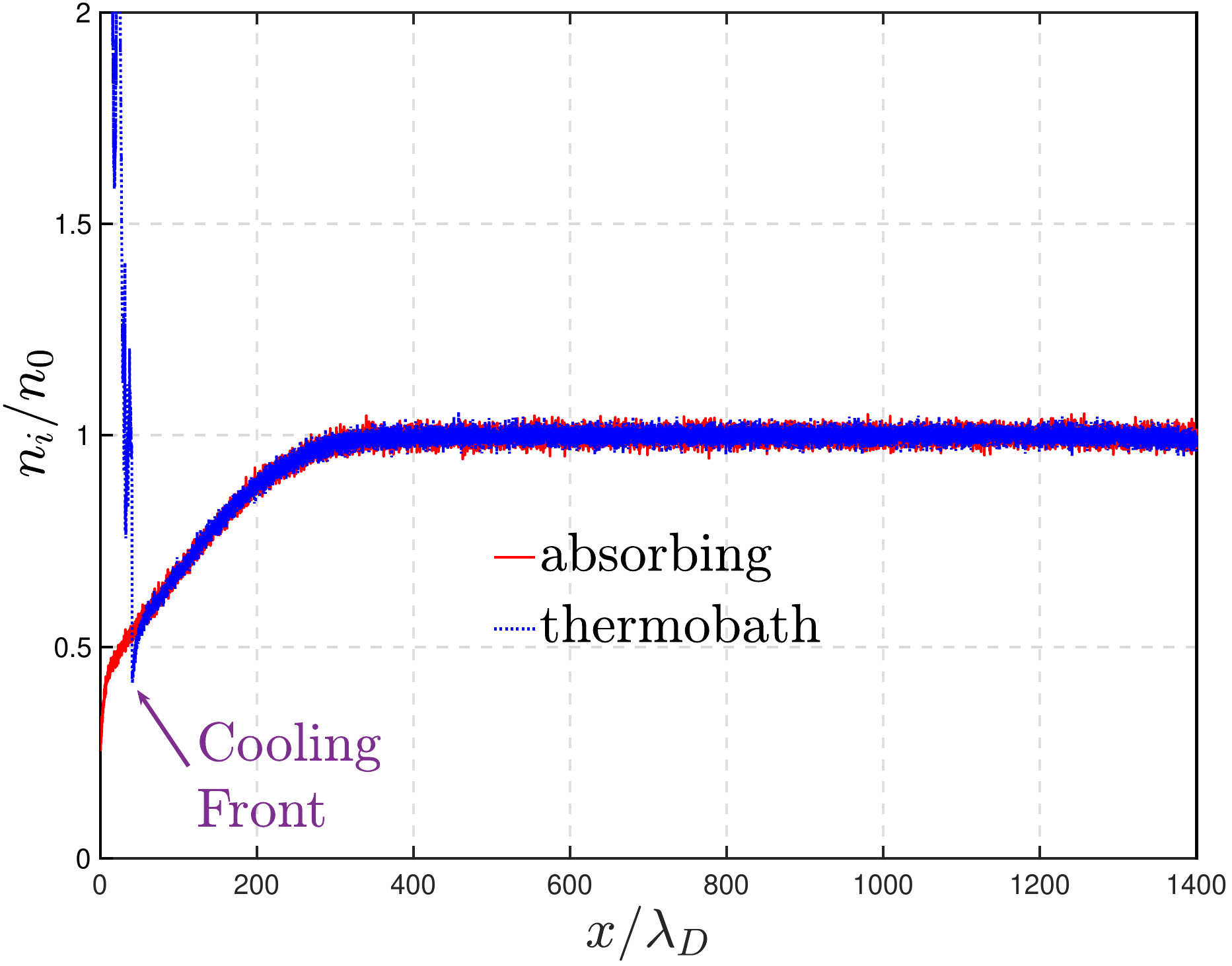}
\includegraphics[width=0.3\textwidth]{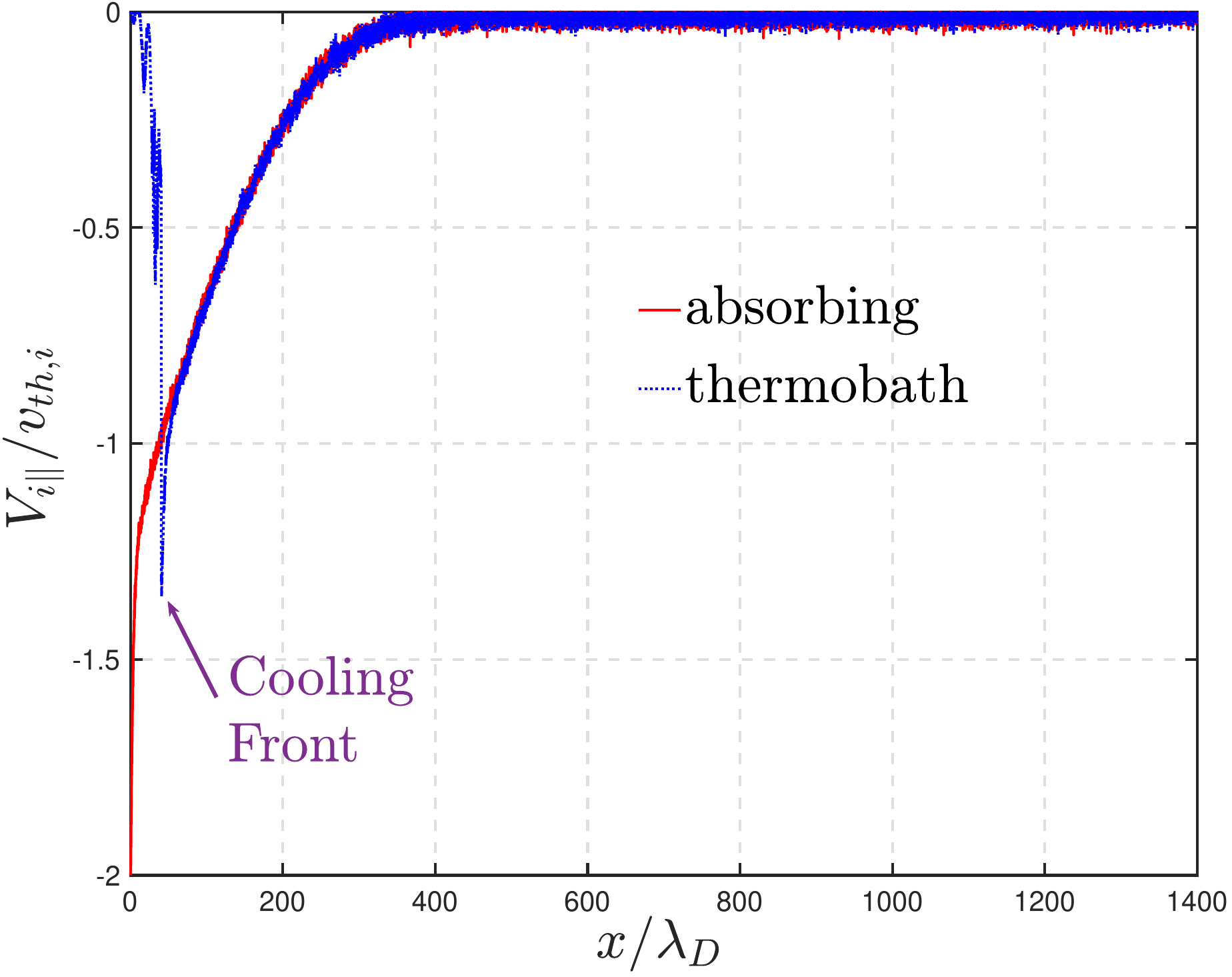}
\includegraphics[width=0.3\textwidth]{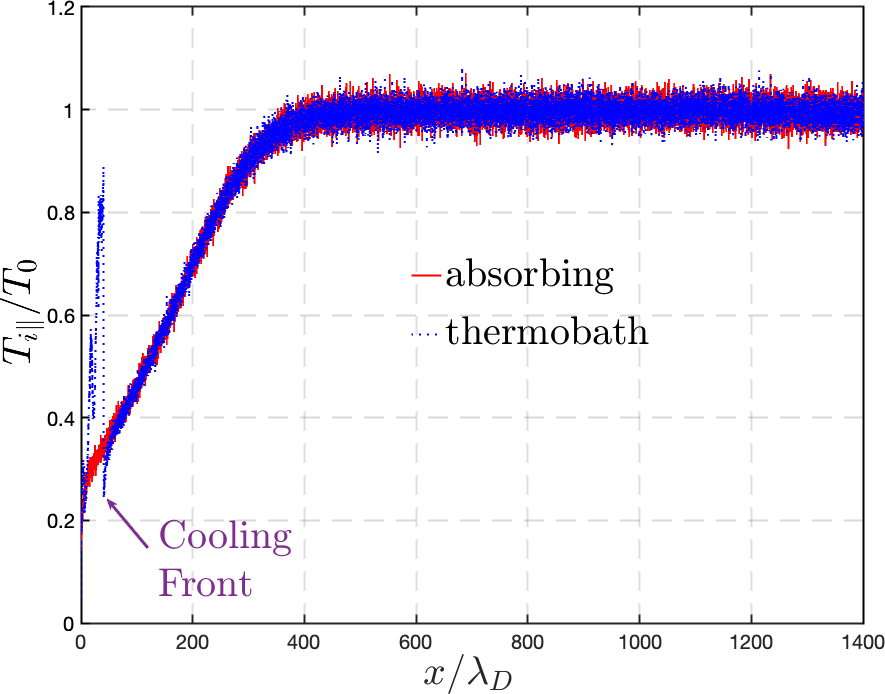}
\caption{Profile of normalized ion density, parallel flow and  temperature at $\omega_{pe}t=1357$ for the simulations in Fig.~\ref{fig:diagram}. The location of the cooling front is marked, ahead of which the ion state variables are aligned with each other for the different boundary conditions. }
\label{fig:ni-vi-Ti-profile}
\end{figure}

To complete these equations, we need closures for the parallel ion
thermal conduction of the parallel degree of freedom $q_{in}\equiv
\int m_i\left(\mathbf{v}_\parallel - V_{i\parallel}\right)^3 f_i
d\mathbf{v}$ and the electron pressure gradient. Within the recession
layer, the ion distribution can be approximated by one-sided cutoff
Maxwellian with a proper shift like in the presheath
region~\cite{tang2016kinetic} and thus we can employ $q_{in}=\sigma_i
T_{i\parallel}\Gamma_{i\parallel}$ with $\sigma_i\sim 1$ the
energy transmission coefficient and $\Gamma_i=n_iV_{i\parallel}$. The
spatial gradient of $q_{en}$ is shown to retain the convective energy
transport scaling $\sim n_eV_{e\parallel}T_{e\parallel}$ over the
recession layer (see the next section) so that we can approximate
$\partial q_{en}/\partial x \approx \sigma_e\partial (n_e
T_{e\parallel} V_{i\parallel})/\partial x$, where $\sigma_e\sim 1$
is analogous to $\sigma_i$. As a result, from the continuity and
energy equations for ions in Eqs.~(\ref{eq-ion-density},
\ref{eq-ion-temperature}) and electrons, which have the same form as
Eqs.~(\ref{eq-ion-density}, \ref{eq-ion-temperature}) but with $e$
replacing $i$ in the subscripts, we obtain
\begin{equation}
\frac{-\xi+(1+\sigma_{e,i})V_{e,i\parallel}}{3+\sigma_{e,i}}\frac{d\textup{ln}p_{e,i\parallel}}{d\xi}=\frac{d\textup{ln}V_{e,i\parallel}}{d \xi}.
\end{equation}
Recalling the quasi-neutral condition $V_{e\parallel}\approx V_{i\parallel}$, one finds
\begin{equation}
\frac{d\textup{ln}p_{e\parallel}}{d\xi}\approx \mu \frac{d\textup{ln}p_{i\parallel}}{d\xi},\label{eq-dpe-dx-Z}
\end{equation}
where $\mu=(3+\sigma_e)/(3+\sigma_i) \times
[-\xi+(1+\sigma_i)V_{i\parallel}]/[-\xi+(1+\sigma_e)V_{i\parallel}]\sim
1$. In physics terms, Eq.~(\ref{eq-dpe-dx-Z}) demonstrates an
universal length scale for $p_{e,i\parallel}$ within the recession
layer since the convective transport scaling dominates the thermal
collapse of $T_{e,i\parallel}$. As a result,
Eqs.~(\ref{eq-ion-density}-\ref{eq-ion-temperature}) form a complete
set of  equations in the form of
\begin{align}
\textbf{A} \left(\frac{d n_i}{d \xi}, \frac{d V_{i\parallel}}{d\xi},\frac{dT_{i\parallel}}{d\xi}\right)^T=0,\label{eq-matrix-ions}
\end{align}
where $\textbf{A}$ is a matrix of $n_i$, $V_{i\parallel}$ and $T_{e,
  i\parallel}$. The non-trivial solution requires $det(\textbf{A})=0$,
yielding
\begin{align}
G&=-\xi^3+\left(3+\sigma_i \right)V_{i\parallel}\xi^2-[\left(3+2\sigma_i\right)V_{i\parallel}^2  -\left(1+\sigma_i/3\right)c_s^2]  \xi \nonumber\\ &+\left(1+\sigma_i\right)V_{i\parallel}^3-\left(1+\sigma_i/3\right)c_s^2V_{i\parallel}=0,\label{eq-U-cubic-Z}
\end{align}
where $c_s=\sqrt{3(\mu ZT_{e\parallel}+T_{i\parallel})/m_i}$ is approximated to the local ion sound speed. A uniquely monotonic solution of $\xi$ can be found in the recession layer as
\begin{equation}
\xi=[\sigma_i^2V_{i\parallel}^2/4+(1+\sigma_i/3)c_s^2]^{1/2}+(1+\sigma_i/2)V_{i\parallel},\label{eq-U-Z}
\end{equation}
by ignoring the explicit dependence on $\xi$ of $\mu$ in $c_s$.
It shows that the recession speed $\xi$ is determined by the local ion flow and sound speed, and thus the recession layer is independent of the boundary condition (e.g., see Fig.~\ref{fig:ni-vi-Ti-profile}). This is not surprising since the cold ions are blocked by the CF, while the recycled electron beam only moderately modifies the electron profiles ahead of the CF.

\begin{figure}[hbt]
\centering
\includegraphics[width=0.32\textwidth]{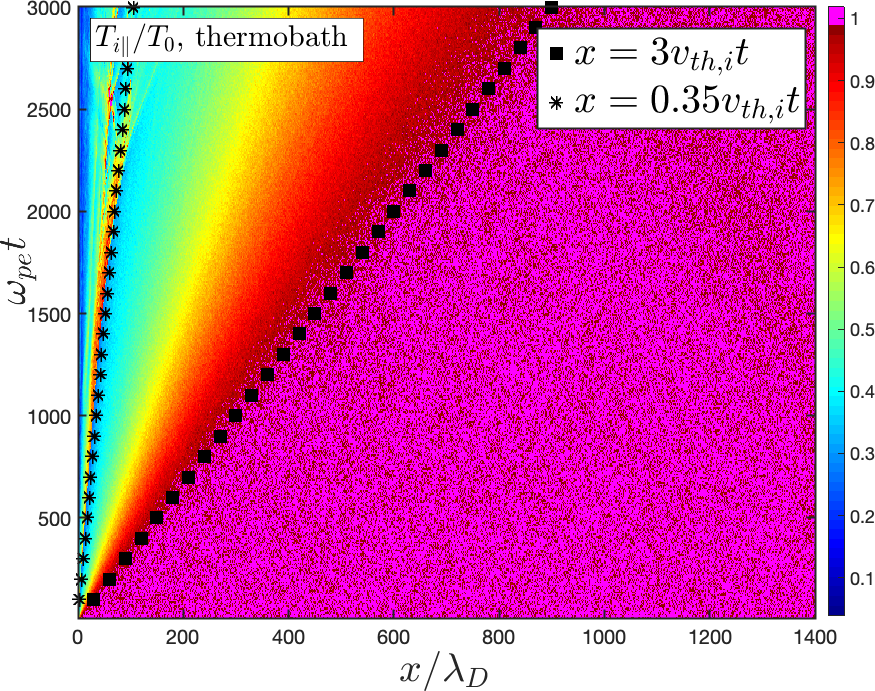}
\includegraphics[width=0.32\textwidth]{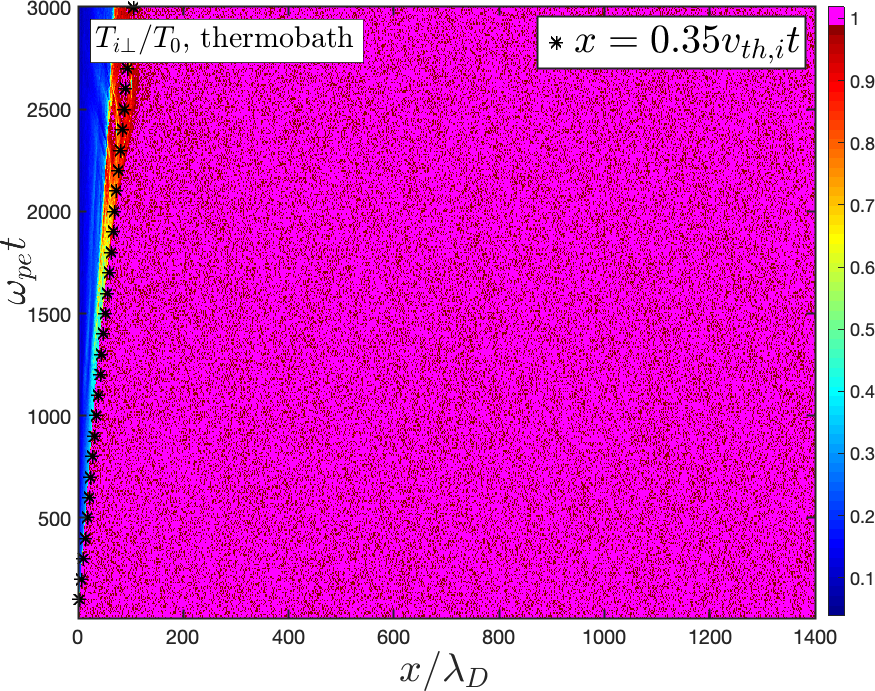}
\includegraphics[width=0.32\textwidth]{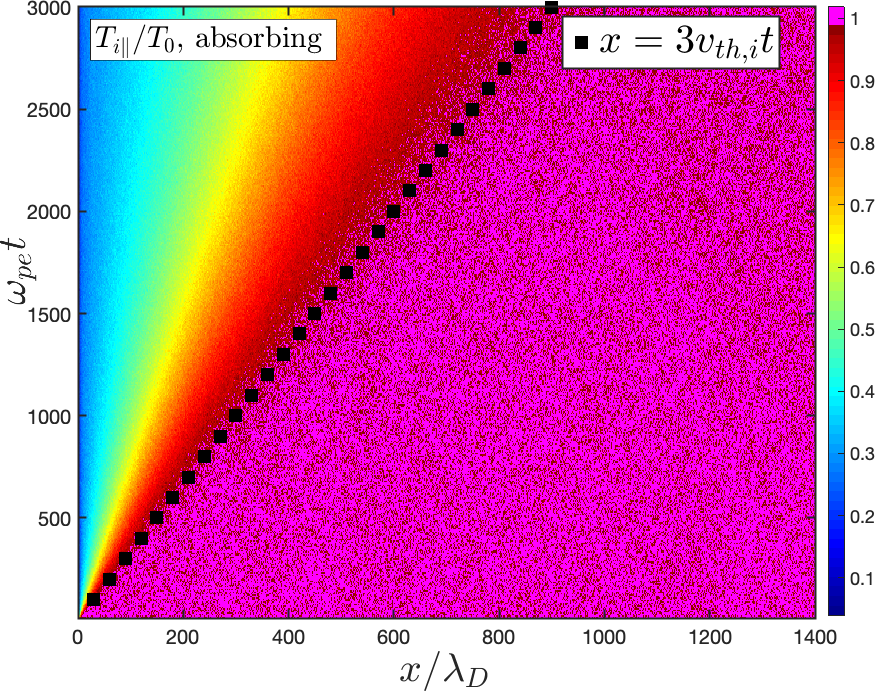}
\caption{Contour plots of normalized ion temperature and the corresponding fitting recession front (black squares) for the same simulations in Fig.~\ref{fig:diagram}, where the left boundary conditions are labeled. For the thermobath boundary case, we also plot the fitting cooling front (black stars), which denotes the onset of $T_{i\perp}$ collapse. The nearly unperturbed values are colored as magenta. }
\label{fig:contour-Ti-reflux}
\end{figure}

Let's first use the self-similar solution of Eq.~(\ref{eq-U-Z}) to
recover a known constraint on the plasma exit flow at an absorbing
boundary where a non-neutral sheath would form next to it. The sheath entrance can not propagate further upstream in this case so
$\xi=0$. Thus Eq.~(\ref{eq-U-Z}) predicts an ion exit flow speed of 
\begin{equation}
V_{i\parallel}=-\sqrt{\frac{1+\sigma_i/3}{1+\sigma_i}}c_s\equiv
-\sqrt{(\beta 3ZT_{e\parallel}+3T_{i\parallel})/m_i},\label{eq-Vi_sh}
\end{equation}
with
\begin{equation}
\beta=\frac{1-\frac{1}{Ze\Gamma_i}\frac{\partial q_{in}}{\partial
    \phi}}{1+\frac{1}{e\Gamma_e}\frac{\partial q_{en}}{\partial
    \phi}},
\end{equation}
and $\Gamma_{e,i}=n_{e,i}V_{e,i\parallel}$. This is consistent with
the Bohm criterion for plasma in steady state ($d/dt=0$) when
including the heat flux in the transport model~\cite{tang2016critical,Yuzhi,li2022transport}. 
%Similar to the Bohm speed at the sheath entrance, for a general position with $\xi\neq0$, we can rewrite the solution in Eq.~(\ref{eq-U-Z}) in the form of
%\begin{equation}
%(-\xi+V_{i\parallel})(-\xi+V_{i\parallel}+\sigma_i V_{i\parallel})=(1+\sigma_i)[(3\beta' ZT_{e\parallel}+3T_{i\parallel})/m_i],\label{eq-ion-flow-U-Z}
%\end{equation}
%with
%\begin{equation}
%\beta'=\frac{1-\frac{2}{Zen_i(2V_{i\parallel}+\xi)}\frac{\partial q_{in}}{\partial \phi}}{1+\frac{2}{en_e(2V_{i\parallel}+\xi)}\frac{\partial q_{en}}{\partial \phi}}\times \frac{1+\sigma_e}{1+\sigma_i}\frac{-\xi+(1+\sigma_i)V_{i\parallel}}{-\xi+(1+\sigma_e)V_{i\parallel}}.
%\end{equation}
More importantly, Eq.~(\ref{eq-U-Z}) predicts the speed of the RF, which is defined as $V_{i\parallel}=0$ 
\begin{align}
U_{RF}=\xi(V_{i\parallel}=0)=\sqrt{(1+\sigma_i/3)}c_s.\label{eq:URF}
\end{align}
Recalling Eqs.~(\ref{eq-ion-density}-\ref{eq-ion-temperature}), such
definition of the RF also denotes the onset of $n_i$ and
$T_{i\parallel}$ drop.  Moreover, since the electron temperature is only
moderately reduced at the RF, $T_{e\parallel}\approx
T_{e\parallel}(U_{PTF})$, we have $U_{RF}\approx 2.8 v_{th,i}$ for
$Z=1$ and $\mu=1$ from Eq.~\eqref{eq:URF}. This is consistent with the
VPIC simulations as shown in Fig.~\ref{fig:contour-Ti-reflux}.

\begin{figure}[hbt]
\centering
\includegraphics[width=0.4\textwidth]{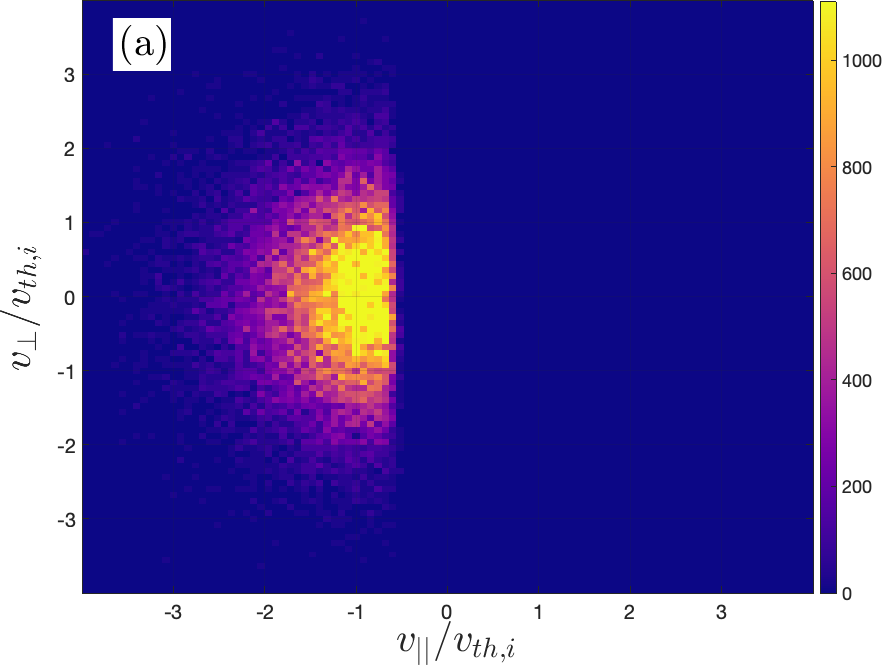}
\includegraphics[width=0.4\textwidth]{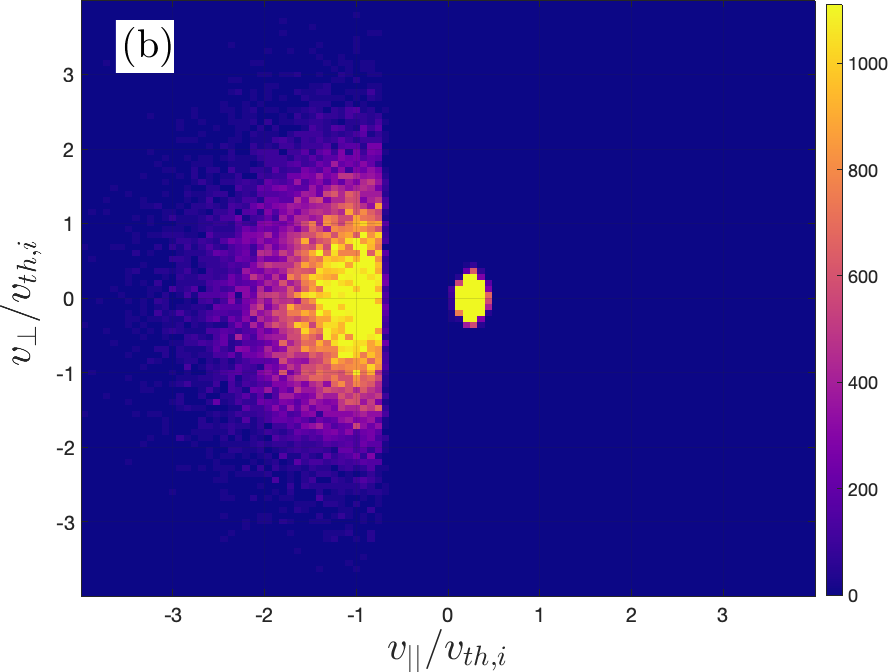}
\caption{Ion distribution just ahead (a) and behind (b) the cooling
  front for the thermobath boundary case corresponding to
  Fig.~\ref{fig:diagram}. Here $v_\parallel$ is the velocity in the
  parallel direction, while $v_\perp$ is the velocity in one of the
  perpendicular directions. Behind the cooling front, there are two
  components in the ion distribution, which have quite different
  perpendicular temperature, indicative of the cooling flow ions from
  the upstream and cold recycled ions from the boundary.}
\label{fig:fi-cooling-front}
\end{figure}

The cooling front (CF) in the thermobath boundary case is where the
hot ions meet the cold ions as illustrated in
Fig.~\ref{fig:fi-cooling-front}. In fact, it is a shock front, across
which all the plasma state variables have jumps as shown in
Figs.~\ref{fig:diagram}-\ref{fig:contour-Ti-reflux}. Particularly, the
ions undergo heating in the parallel direction when the plasma flow
runs into the CF, where the substantial ion flow energy in the
recession layer is converted into ion thermal energy as shown in
Figs.~\ref{fig:diagram}, \ref{fig:ni-vi-Ti-profile}, and
\ref{fig:contour-Ti-reflux}. Such conversion is via the mixing of the
cooling flow ions with the cold recycled ions (e.g., see
Fig.~\ref{fig:fi-cooling-front}), the latter of which are accelerated
from the boundary to the CF by the inverse pressure gradient due to
the pressure pile up at the boundary. As a result, these cold recycled
ions will offset the plasma flow generated by the surrounding ions as
shown in Fig.~\ref{fig:ni-vi-Ti-profile}. In sharp contrast to ions,
the electrons will undergo cooling via dilution with high-density cold
electrons so $T_{e\parallel}\sim T_w$ behind the CF as shown in
Figs.~\ref{fig:diagram} and \ref{fig:contour-Te-reflux}. Moreover, the
presence of the CF and the associated cooling zone behind the CF is of
fundamental importance to $T_{i\perp}$ and $T_{e\perp}$ cooling via
dilution and thus the CF represents a deep cooling of electrons.

To find the speed of the shock (cooling) front, we can match the
conserved quantities across the front while simply ignoring the thermal
conduction flux due to the convective scaling of its gradient. In the
moving frame of reference with the shock front, we have
\begin{align}
\label{eq-ion-density-conserv}
& n_{i1} \tilde{V}_{i\parallel1} = n_{i2} \tilde{V}_{i\parallel2}, \\
  & m_in_{i1}\tilde{V}_{i\parallel1}^2+(T_{i\parallel 1}+ZT_{e\parallel 1})n_{i1}= m_in_{i2}\tilde{V}_{i\parallel2}^2+(T_{i\parallel2}+ZT_{e\parallel2})n_{i2},\label{eq-ion-momentum-conserv} \\
&  m_in_{i1}\tilde{V}_{i\parallel 1}^3+3(T_{i\parallel1}+ZT_{e\parallel1})n_{i1}\tilde{V}_{i\parallel1}=m_in_{i2}\tilde{V}_{i\parallel2}^3+3(T_{i\parallel2}+ZT_{e\parallel2})n_{i2}\tilde{V}_{i\parallel},\label{eq-ion-temperature-conserv}
\end{align}
where the subscripts $1$ and $2$ denote, respectively, the ion
variables behind (downstream) and ahead (upstream) of the shock front,
and $\tilde{V}_{i\parallel}=V_{i\parallel}-U_{CF}$ with $U_{CF}$ being
the CF speed. As a result, we find
\begin{equation}
\frac{\tilde{V}_{i\parallel1}}{\tilde{V}_{i\parallel2}}=\frac{\tilde{V}_{i\parallel2}^2+3(ZT_{e\parallel2}+T_{i\parallel2})/m_i}{2\tilde{V}_{i\parallel2}^2}\equiv\frac{1+1/M_2^2}{2}.\label{eq-jump-of-vi}
\end{equation}
In the upstream (recession layer), we have shown that
$-V_{i\parallel}+\xi=c_s=\sqrt{3(ZT_{e\parallel}+T_{i\parallel})/m_i}$
as seen from Eq.~(\ref{eq-U-Z}) with $\sigma_i=0$. Therefore, the Mach
number $M_2= |V_{i\parallel}-U_{CF}|/c_s$ is near unity and thus the
CF is a low Mach number shock. It should be noted that, near the CF,
the ion flow is further accelerated by the large ambipolar field
compared to that for the absorbing boundary (e.g., see
Fig.~\ref{fig:ni-vi-Ti-profile}), increasing $M_2$ slightly above the
unity. These conditions ensure the stable flow upstream and downstream
of the shock~\cite{kuznetsov2018parallel}. Further simplification to
obtain the CF speed will be $|V_{i\parallel1}|\ll U_{CF}$ due to the
offset of plasma velocity by the cold recycled particles as shown in
Fig.~\ref{fig:ni-vi-Ti-profile} so that
\begin{equation} 
U_{CF} \approx -\tilde{V}_{i\parallel2} \approx c_{s,2},
\end{equation} 
with $M_2\approx 1$. This indicates that the CF propagates with the upstream ion sound speed. Since the plasma temperature at the CF is considerably 
lower than that at the RF, we have $U_{CF} < U_{RF}.$
%More detailed physics underlying the cooling/shock front will be presented somewhere else.

\iffalse
 It is noteworthy that $U_{CF}$ positively depends on $T_w$ as shown in Fig.~\ref{fig:cooling-front-speed-Tw}. This is because smaller $T_w$ will require more aggregated particles to pile up the pressure, and thus the plasma flow towards the cooling spot at the CF should be larger. A detailed study of the shock front is beyond the scope of this paper.

\begin{figure}[hbt]
\centering
\includegraphics[width=0.65\textwidth]{figures/cooling_front_speed_new.pdf}
\caption{Fitted speeds of cooling front versus $T_w$ from the simulations of collisionless plasma.  }
\label{fig:cooling-front-speed-Tw}
\end{figure}

\fi

\section{\label{sec:heatflux} Electron thermal conduction flux under the ambipolar transport constraint}

In this section, we investigate the electron thermal conduction flux
within the recession layer under the ambipolar transport constraint
$V_{e\parallel}\approx V_{i\parallel}$. Both model analyses and
first-principle simulations using VPIC will be provided.

\subsection{Analytical results}
 For the electron distribution in Eq.~(\ref{eq:truncated-fe}), the electron density, parallel flow, temperature, and thermal conduction heat flux are given in Eqs.~(\ref{eq:ne-coldbeam}-\ref{eq-qen}), from which one finds an expression for $T_{e\parallel}$ and $q_{en}$
\begin{align}
  T_{e\parallel} =&
  T_0 \left[
    1 - \frac{n_b}{n_e} + \frac{V_{e\parallel} v_c}{v_{th,e}^2} - \frac{V_{e\parallel}^2}{v_{th,e}^2}
    \right], \label{eq:Tepara-coldbeam-new}\\
    q_{en} 
  =&-2n_bv_cT_0+ \left(\frac{v_c^2}{v_{th,e}^2}+2\right)n_eV_{e\parallel}T_0
  - 3n_e T_{e\parallel} V_{e\parallel} - n_em_eV_{e\parallel}^3.\label{eq:qen-coldbeam}
\end{align}
Then from Eqs.~(\ref{eq:nV-cold-beam}, \ref{eq:Tepara-coldbeam-new}, \ref{eq:qen-coldbeam}), we see that the impact of the cold beam can be ignored in the limit of a small cold beam component
\begin{align}
n_b v_c < n_e V_{e\parallel}.
\end{align}
Since $v_c \sim v_{th,e},$ one has in this limit
\begin{align}
\frac{n_b}{n_e} < \sqrt{\frac{m_e}{m_i}},
\end{align}
under which we cover the absorbing boundary case with $n_b=0$. Otherwise, the cold beam can make a big difference.

For an absorbing boundary or a small beam component, the ambipolar transport will limit the cutoff velocity to be large $v_c \propto
v_{th,e} \sqrt{2\ln (v_{th,e}/V_{i\parallel})},$ so that
$V_{e\parallel}\approx V_{i\parallel}$ as seen from Eq.~\eqref{eq:nV-cold-beam}. However, in the large cold beam component limit, 
\begin{align}
n_b v_c \sim n_b v_{th,e} \gg n_e V_{e\parallel},
\end{align}
or equivalently, $n_b /n_e\gg \sqrt{m_e/m_i}$, one must have, to the leading order in $\sqrt{m_e/m_i},$
\begin{align}
n_b v_c \approx \frac{n_mv_{th,e}}{\sqrt{2\pi}} e^{-v_c^2/2v_{th,e}^2}, \label{eq:cold-e-beam-flux}
\end{align}
which means that the cold beam and the truncated Maxwellian both carry
significant electron particle fluxes, but nearly cancel each other to
produce a much slower electron flow $V_{e\parallel}$ that matches onto
$V_{i\parallel}$ for ambipolar transport. Compared to the small beam
component limit, the requirement for large $v_c$ and hence $\Delta
\Phi$ in the absorbing boundary case is relaxed due to the large
electron beam component. Therefore, the reflecting potential and hence
$U_{PTF}$ for the thermobath boundary is smaller than those for the
absorbing boundary as discussed in section
\ref{sec:electronfronts}. From Eqs.~(\ref{eq:ne-coldbeam},
\ref{eq:cold-e-beam-flux}), we obtain
\begin{align}
\frac{n_b}{n_e} = \frac{1}{1 + \sqrt{\frac{\pi}{2}}\left[1+\textup{Erf}(v_c/\sqrt{2}v_{th,e})\right] \frac{v_c}{v_{th,e}} e^{v_c^2/2v_{th,e}^2}}.
\end{align}
One can easily verify that $n_b/n_e$ is a monotonically decreasing function of $v_c$. It is important to note that $v_c$ can not vanish (or $n_b=n_e$), otherwise, $T_{e\parallel}$ would become negative as seen from Eq.~(\ref{eq:Tepara-coldbeam-new}). However, $v_c$ can become very small and so does $T_{e\parallel}$, indicating that the
plasma temperature can drop to the value set by the temperature of
the beam-like cold plasma at the CF. 

The necessary constraint to sustain the large gradient of $T_{e\parallel}$ within the recession layer to drive the plasma flow towards the cooling spot is on the spatial gradient of $q_{en}$, which can be argued in the following. Assuming the recession layer span a length of $L_R$, from Eq.~(\ref{eq-electron-energy}), the convective energy transport terms (the term that is proportional to $V_{e\parallel}$ itself or its gradient)
follow the scaling of $n_e T_{e\parallel} V_{e\parallel}/L_R.$ Therefore, a necessary
condition for forming the recession layer by keeping a large $T_{e\parallel}$ gradient will require
\begin{align}
\frac{\partial q_{en}}{\partial x} \lesssim \frac{n_e T_{e\parallel}
V_{e\parallel}}{L_R}. \label{eq:qen-x-gradient}
\end{align}
This is indeed the case due to the ambipolar transport constraint. To see it, we re-write $q_{en}$ in Eq.~(\ref{eq:qen-coldbeam}) as
\begin{align}
  q_{en} \approx - \alpha_e n_e v_{th,e} T_0
  + \sigma_e n_e V_{e\parallel} T_{e\parallel} ,\label{qen-recession-layer-coldbeam}
\end{align}
where
\begin{align}
  \alpha_e & \equiv 2 \frac{n_b}{n_e}\frac{v_c}{v_{th,e}}, \\
  \sigma_e & \equiv \left(\frac{v_c^2}{v_{th,e}^2}+2\right)\frac{T_0}{T_{e\parallel}} -3.
\end{align}
The second term carries over from the absorbing boundary case where no
cold electron beam component is present, while the first term is
entirely due to the presence of a cold beam component, as it is
proportional to $n_b/n_e$. This first term is of great importance
since it covers the conventional electron free-streaming scaling of
$q_{en}$, although the coefficient critically depends on the
fractional density of the cold beam component.

In the case of an absorbing boundary, $n_b/n_e=0,$ we have
\begin{align}
q_{en} \approx \sigma_e n_e V_{e\parallel} T_{e\parallel},\label{eq:qen-trap-elec}
\end{align}
which illustrates that the parallel electron heat flux itself scales as the convective energy transport scaling. While in the large cold beam limit, the first term of $q_{en}$ dominates
\begin{align}
  q_{en} \approx - \alpha_e n_e v_{th,e} T_0,\label{eq:qen-free-streaming}
\end{align}
for $v_c>V_{e\parallel}$ so that $q_{en}$ itself has electron mass scaling $m_e^{-1/2}$, recovering the free-streaming limit. However, its
spatial gradient over the recession layer retains the convective
transport scaling. This is because the cold beam continuity equation in the collisionless recession layer has
\begin{equation}
    n_bv_c\approx const.,\label{eq-beam-flux}
\end{equation}
to obtain which we have ignored $\partial n_b/\partial t$ in the beam continuity equation because if we seek the self-similar solution between the ion fronts, we have $\partial n_b/\partial t \approx -\xi\partial n_b/\partial x\ll v_c\partial n_b/\partial x$ with $\xi\sim c_s\ll v_c\sim v_{th,e}$. This indicates that $\partial(n_bv_c)/\partial x=-\partial n_b/\partial t \sim c_s n_b/L_R$ contributes less to $\partial q_{en}/\partial x$ than that from the second term in Eq.~(\ref{qen-recession-layer-coldbeam}) since $n_b< n_e$. Therefore, we have reached
the interesting point that the electron parallel heat flux is
primarily carried by the electron beam to follow the free electron
streaming scaling, but remarkably, its spatial gradient is given by
the convective energy transport scaling. 

\subsection{First-principles kinetic simulations}
The first-principles kinetic simulations have confirmed all the above
analytical results. Specifically, to check the convective transport
scaling of $q_{en}$, i.e., $q_{en}\propto m_i^{-1/2}$, we have
conducted simulations by employing different ion-electron mass ratios,
i.e., $m_i/m_e=100,~400,~900$ and $1600$. Eq.~(\ref{eq-U-Z}) indicates
that both the ion flow and local recession speed $\xi$ in the
recession layer scales with $m_i^{-1/2}$. Therefore, to overlap the
recession layer to the same location for different $m_i$, we should
choose temporal moments with constant $\omega_{pi}t$ for different cases. Here
$\omega_{pi}$ is the ion plasma frequency.

\begin{figure}[hbt]
\centering
\includegraphics[width=0.45\textwidth]{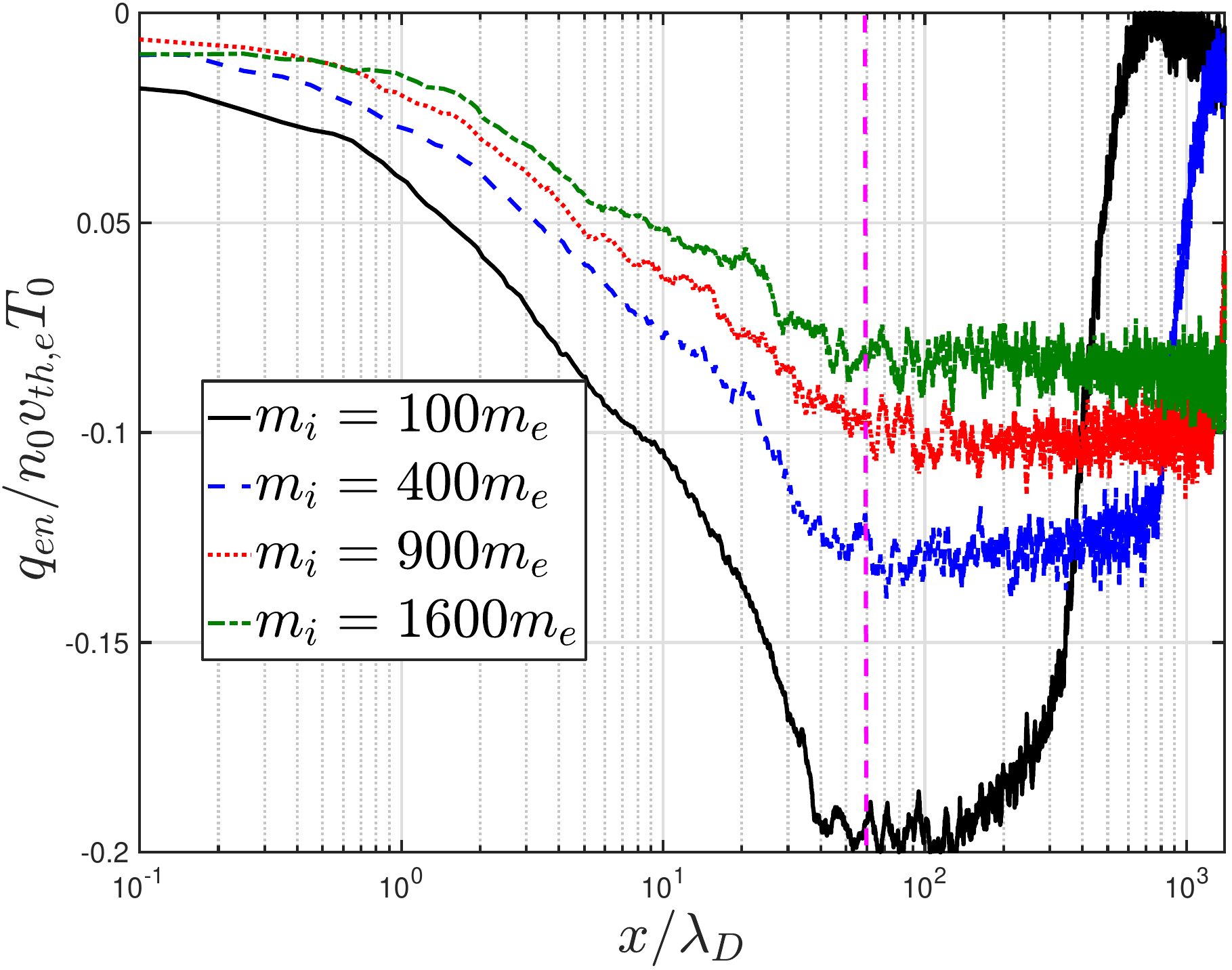}
\includegraphics[width=0.45\textwidth]{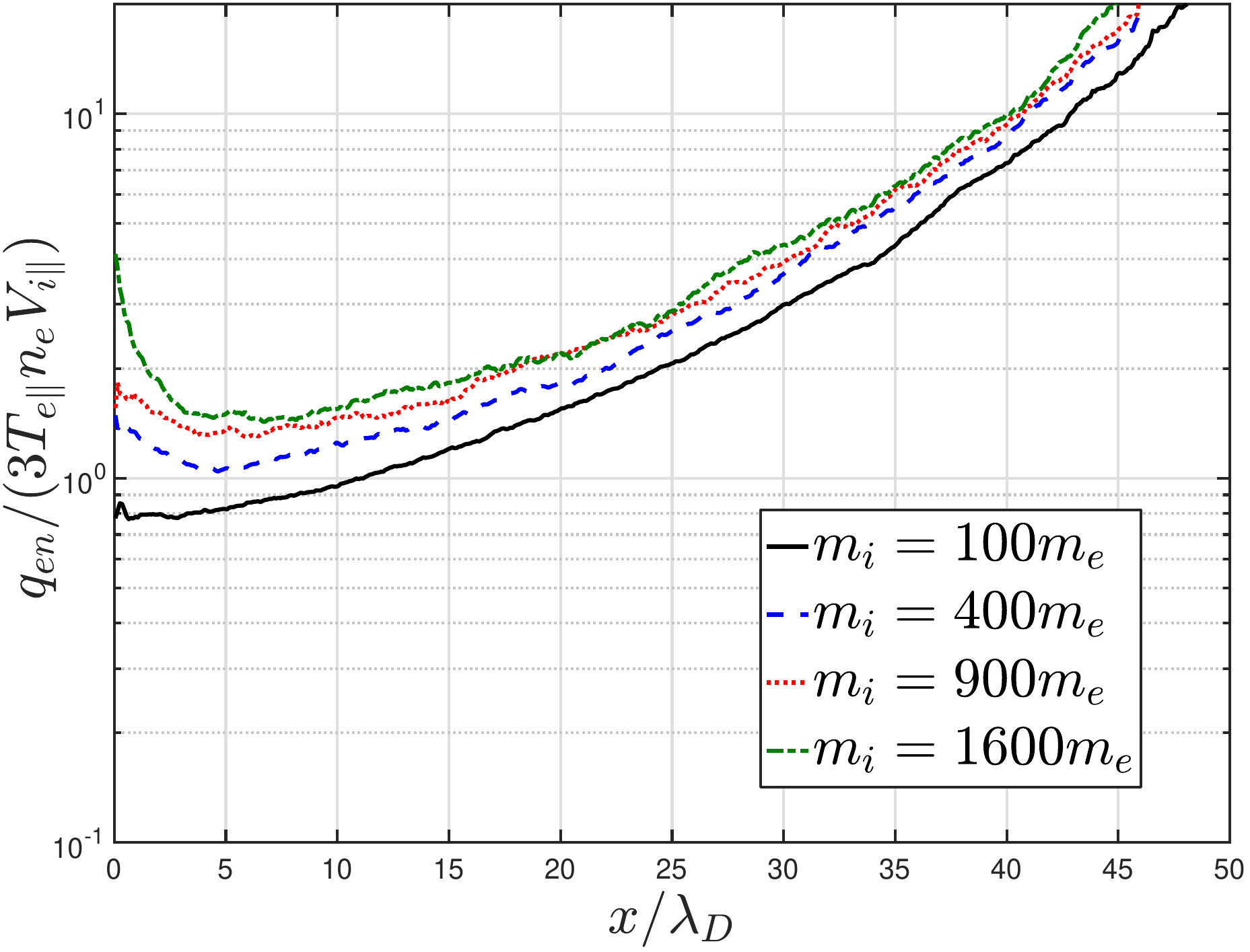}
\caption{Profiles of $q_{en}$ at $\omega_{pi}t=20.3$ for different $m_i/m_e$ with the absorbing boundary. Left: $q_{en}$, normalized by the initial quantities $n_0T_0v_{th,e}$, at the whole domain, where the recession front at $x\approx 60\lambda_D$ is labeled by the vertical dashed line; Right: $q_{en}$, normalized by the convection heat flux $3n_eV_{i\parallel}T_{e\parallel}$, behind the recession front (away from the recession front where $V_{i\parallel}\approx 0$). }
\label{fig:heatfluxes-recession-absrobing}
\end{figure}

In Fig.~\ref{fig:heatfluxes-recession-absrobing} we plot the profiles
of $q_{en}$ for the absorbing boundary case, which demonstrates that
$q_{en}$ itself indeed follows the convective transport scaling
$q_{en}\propto V_{i\parallel}$ within the recession layer, although
the coefficient $\sigma_e$ slightly increases with $m_i$. Such
variation of $\sigma_e$ comes from the weakly positive dependence of
$v_c$ on $m_i$ via $v_c\sim
v_{th,e}\sqrt{2\ln(v_{th,e}/V_{i\parallel})}$.

\begin{figure}[hbt]
\centering
\includegraphics[width=0.45\textwidth]{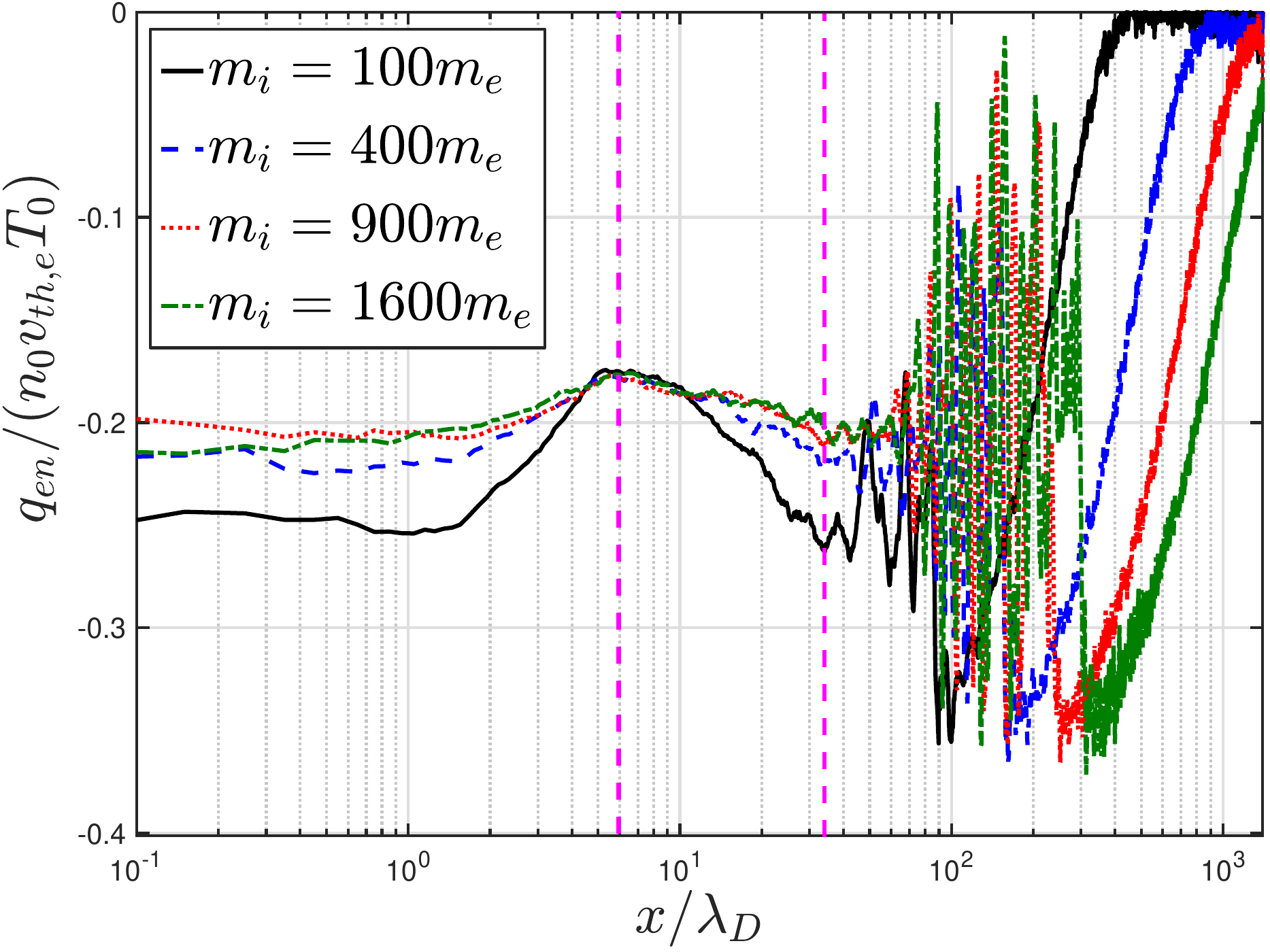} 
\includegraphics[width=0.45\textwidth]{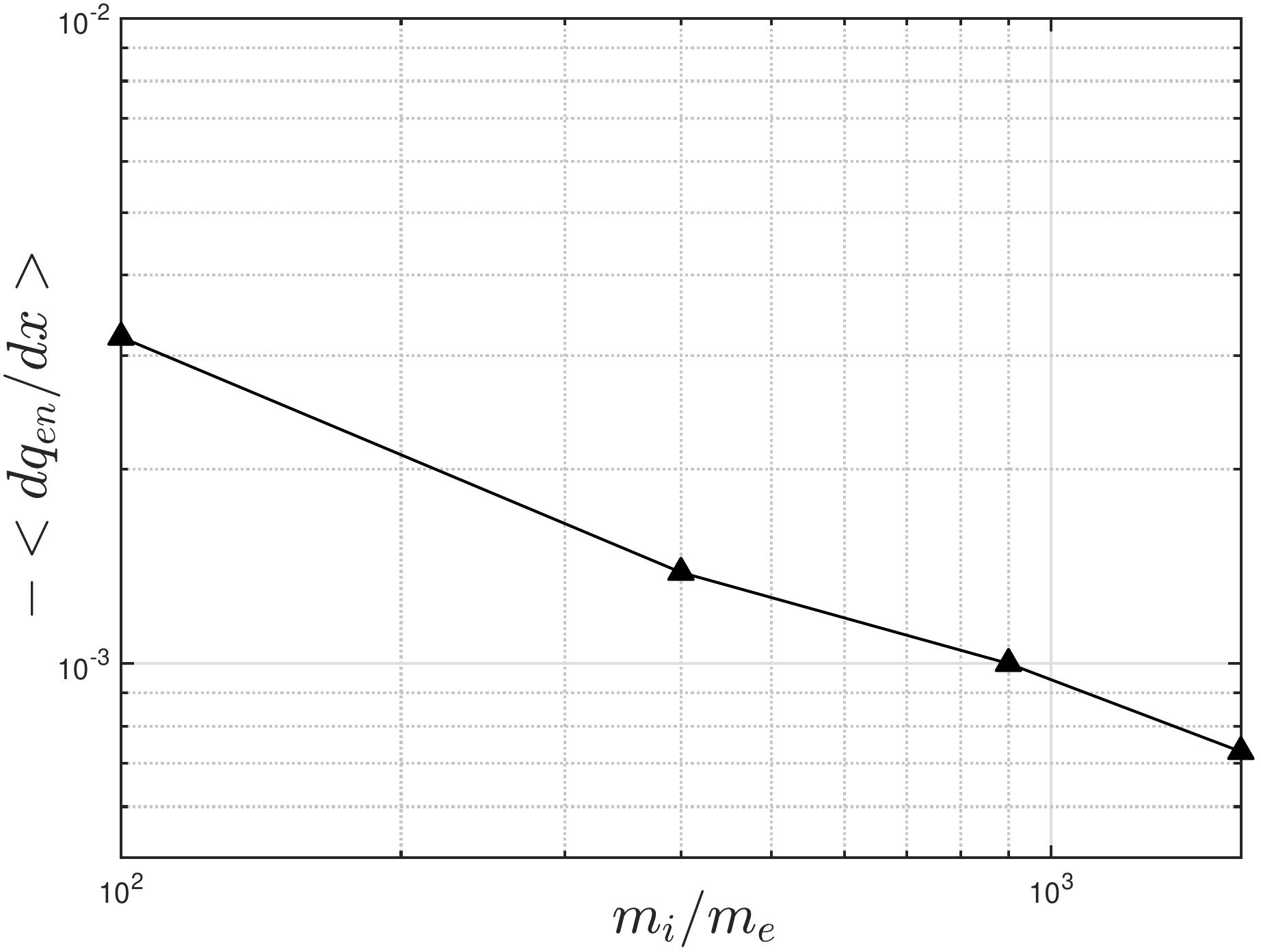}
\caption{Profiles of $q_{en}$ (left) and its averaged gradient
  $<dq_{en}/dx>$ (right) in the recession layer at $\omega_{pi}t=13.6$
  for the thermobath boundary with $T_w=0.01$. The averaged region in
  the recession layer is labeled by vertical dashed lines.}
\label{fig:heatfluxes-recession-layer-diff-mi}
\end{figure}

\begin{figure}[h]
\centering
\includegraphics[width=0.6\textwidth]{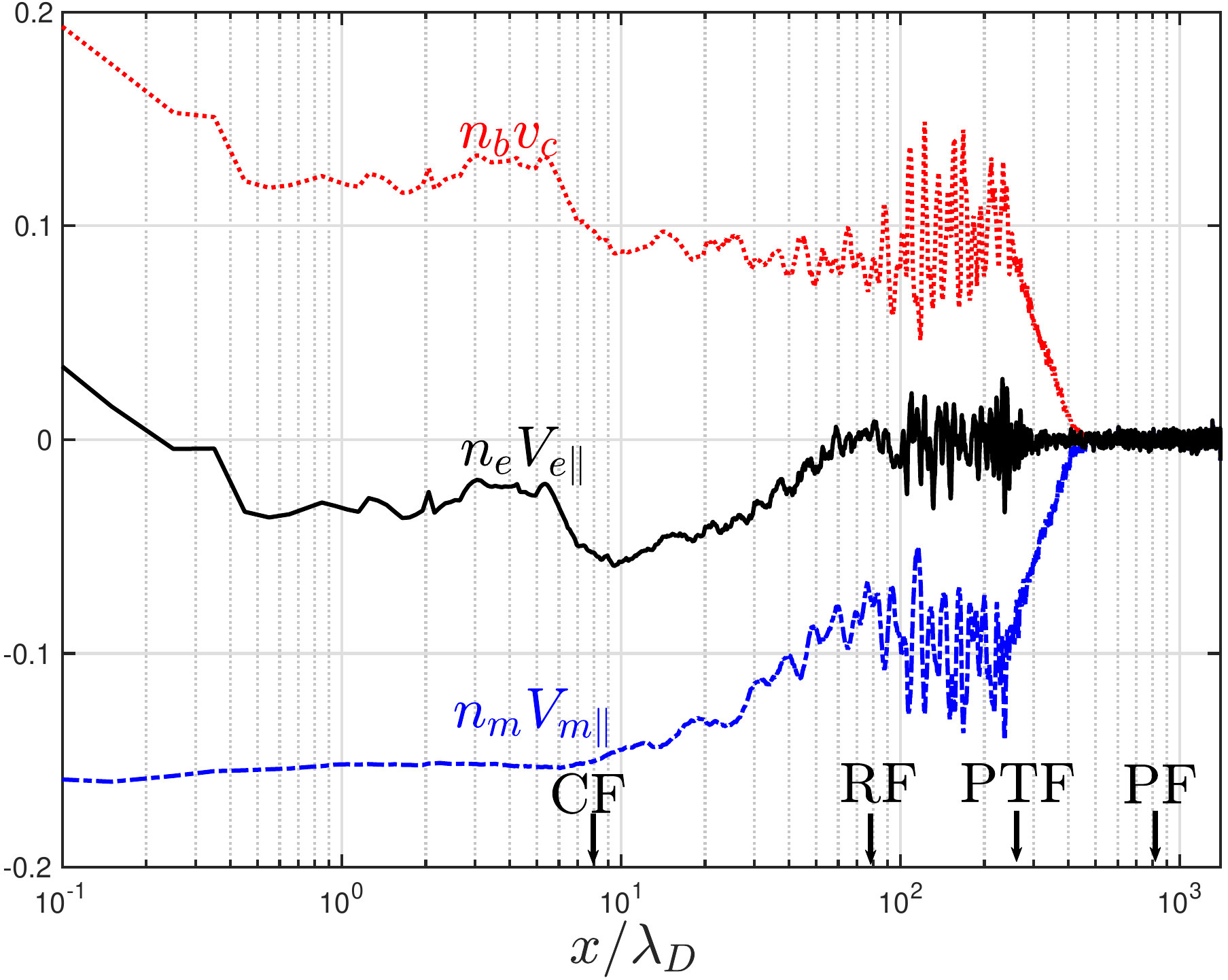}
\caption{$n_eV_{e\parallel}$, $n_bv_c$ and $n_mV_{m\parallel}\equiv
  n_eV_{e\parallel}-n_bv_c$, in the unit of $n_0v_{th,e}$, at
  $\omega_{pe}t=271$ for the thermobath boundary with $m_i/m_e=100$
  and $T_w=0.01$. The locations for the four fronts are labeled.}
\label{fig:neVenbvc}
\end{figure}

For the case of a thermobath boundary,
Fig.~\ref{fig:heatfluxes-recession-layer-diff-mi} shows that $q_{en}$
is nearly independent of $m_i$ and instead recovers a flux-limiting
form for the heat conduction flux with $\alpha_e\sim 0.2$ in the recession layer. But its
gradient within the recession layer still has the convective transport
scaling, i.e., $<dq_{en}/dx>\propto m_i^{-1/2}$. As discussed in
Eq.~(\ref{qen-recession-layer-coldbeam}), the free-streaming form
is recovered for $q_{en}$ because of the dominating cold beam contribution of
$-2n_bv_cT_0.$ This cold beam term, however, doesn't contribute to $<dq_{en}/dx>$
since the cold beam flux $n_bv_c$ in the collisionless recession layer
is constant as shown in Eq.~(\ref{eq-beam-flux}). To quantify the beam
contribution to the electron heat flux as well as the particle flux,
we have separated the cold recycled electrons from the original
electrons. In Fig.~\ref{fig:neVenbvc}, we show the contributions to
the electron particle flux $n_eV_{e\parallel}$ from both recycled
$n_bv_c$ and the original trapped electrons $n_mV_{m\parallel}\equiv
n_eV_{e\parallel}-n_bv_c$ for the $m_i=100m_e$ case. It shows that
both the recycled and the original trapped electrons carry significant
electron particle flux but nearly cancel each other to generate much
smaller $n_eV_{e\parallel}\ll n_bv_c$. Notice that $n_bv_c$ has the
electron scaling $n_bv_c\propto v_{th,e}$ while $n_eV_{e\parallel}$
has the ion scaling $n_eV_{e\parallel}\propto v_{th,i}$ so that the
difference between them is even stronger for larger
$m_i$. Fig.~\ref{fig:neVenbvc} also confirms that $n_bv_c$ is nearly
constant within the recession layer as expected, so it doesn't
contribute to $dq_{en}/dx$. It is worth noting that the cold electron
front is ahead of the PTF. Ahead of the PTF, $n_bv_c$
varies in space so that $q_{en}$ itself and its gradient continuously
follow the free-streaming scaling.

\section{\label{sec:conclusion}Conclusions}

In conclusion, the thermal collapse of a nearly collisionless plasma
interacting with a cooling spot is investigated both theoretically and
numerically. For both types of cooling spots that signify,
respectively, the radiative cooling masses (thermobath boundary) and
a perfect particle and energy sink (absorbing boundary), we found that the
thermal quench comes about in the form of propagating fronts that
originate from the cooling spot. Particularly, for the thermobath
boundary, two fast electron fronts have speeds that scale with the
electron thermal speed $v_{th,e}$, and two slow ion fronts propagates
at local ion sound speed $c_s$. The former denotes the fast but
moderate cooling of electrons, while the latter represents the slow
but aggressive cooling of electrons and ions.

The underlying physics behind these propagating fronts have been
investigated, in which the electron thermal conduction heat flux
$q_{en}$ is found to play an essential role. Specifically, the electron fronts
are completely driven by $q_{en},$ which 
follows the free-streaming
form ($q_{en} \propto n_e v_{th,e} T_{e\parallel}$). Such large
thermal conduction flux is reminiscent of a very limited amount of
$T_{e\parallel}$ drop over a very large volume. In contrast, the ion fronts
are formed as a result of the full transport physics, the crucial one
of which is the ambipolar transport constraint. Due to such ambipolar transport
constraint, $q_{en}$ in the recession layer itself follows the convective energy transport
scaling with the parallel plasma flow $V_{i\parallel}$ for the
absorbing boundary. While for the thermobath boundary, the cold
electron beam will restore the free-streaming limit of $q_{en}\sim
-2n_bv_cT_0\propto v_{th,e}$, but its spatial gradient will follow the
convective transport scaling since the beam particle flux ($n_bv_c$)
remains nearly constant within the recession layer.  As a result, the
electron temperature and hence the pressure retains large spatial
gradient to drive the plasma flow toward the cooling spot.

For a thermobath boundary that provides an energy sink while
re-supplying cold particles, the plasma cooling flow eventually
terminates against the cooling spot via a plasma shock, which we named
the cooling front since it signifies the deep cooling of electrons to
the radiatively clamped temperature $T_w$. It is shown that such a
shock front will convert the ion flow energy into the ion thermal
energy via the mixing of hot and cold ions behind the front, which
completely blocks the cold ions from migrating upstream. Therefore,
the cooling front and the associated cooling zone behind the cooling
front are of fundamental importance to $T_{i\perp}$ and $T_{e\perp}$
cooling via dilution. Unlike the recycled cold ions, part of the
recycled electrons can penetrate through the cooling front to reach
the electron fronts, causing further cooling of the electrons behind
the electron fronts.

\section*{ACKNOWLEDGMENTS}
We thank the U.S. Department of Energy Office of Fusion Energy
Sciences and Office of Advanced Scientific Computing Research
for support under the Tokamak Disruption Simulation (TDS) Scientific
Discovery through Advanced Computing (SciDAC) project, and the Base
Theory Program, both at Los Alamos National Laboratory (LANL) under
contract No. 89233218CNA000001. This research used
resources of the National Energy Research Scientific Computing Center
(NERSC), a U.S. Department of Energy Office of Science User Facility
operated under Contract No. DE-AC02-05CH11231 and the LANL Institutional Computing Program, which is
supported by the U.S. Department of Energy National Nuclear Security
Administration under Contract No. 89233218CNA000001.

\bibliography{reference_fronts.bib}% Produces the bibliography via BibTeX.

\end{document}